\documentclass[letterpaper,twocolumn,10pt]{article}
\usepackage{usenix-2020-09}

\DeclareFixedFont{\ttb}{T1}{txtt}{bx}{n}{8} 
\DeclareFixedFont{\ttm}{T1}{txtt}{m}{n}{8}  

\usepackage{color}

\definecolor{deepblue}{rgb}{0,0,0.5}
\definecolor{deepred}{rgb}{0.6,0,0}
\definecolor{deepgreen}{rgb}{0,0.5,0}
\definecolor{applegreen}{rgb}{0.35, 0.61, 0.0}
\definecolor{harvardcrimson}{rgb}{0.69, 0.0, 0.09}

\mathchardef\hyphenmathcode=\mathcode`\-

\usepackage{listings}

\let\origlstlisting=\lstlisting
\let\endoriglstlisting=\endlstlisting
\renewenvironment{lstlisting}
    {\mathcode`\-=\hyphenmathcode
     \everymath{}\mathsurround=0pt\origlstlisting}
    {\endoriglstlisting}

\everymath{}

\usepackage{ifthen}

\newif\ifcomments
\commentsfalse
\ifthenelse{\boolean{comments}}{
\providecommand{\todo}[1]{{\protect\color{red}{\bf [TODO: #1]}}}
\providecommand{\melih}[1]{{\protect\color{teal}{\bf [Melih: #1]}}}
\providecommand{\samyu}[1]{{\protect\color{green}{\bf [Samyu: #1]}}}
\providecommand{\vinamra}[1]{{\protect\color{pink}{\bf [Vinamra: #1]}}}
\providecommand{\alvin}[1]{{\protect\color{blue}{\bf [Alvin: #1]}}}
\providecommand{\ion}[1]{{\protect\color{orange}{\bf [Ion: #1]}}}
\providecommand{\mike}[1]{{\protect\color{purple}{\bf [Mike: #1]}}}
\providecommand{\lianmin}[1]{{\protect\color{brown}{\bf [Lianmin: #1]}}}
\providecommand{\suresh}[1]{{\protect\color{yellow}{\bf [Suresh: #1]}}}
}{
\providecommand{\todo}[1]{}
\providecommand{\melih}[1]{}
\providecommand{\samyu}[1]{}
\providecommand{\vinamra}[1]{}
\providecommand{\alvin}[1]{}
\providecommand{\ion}[1]{}
\providecommand{\mike}[1]{}
\providecommand{\lianmin}[1]{}
\providecommand{\suresh}[1]{}
}

\usepackage{authblk}
\usepackage{mathtools}
\usepackage{xcolor}
\usepackage{graphicx}
\usepackage{subcaption}
\usepackage{caption}
\usepackage{float}

\usepackage{amsmath,amsfonts}
\usepackage{syntax}
\usepackage{mathpartir}
\usepackage{xspace}
\usepackage{soul}
\usepackage[ruled,vlined]{algorithm2e}
\usepackage{wrapfig}

\newcommand{\seq}[2]{ #1 \, ; #2}

\newcommand{\R}{\textbf{R}}

\renewcommand{\a}{\textbf{a}}

\newcommand{\tensordot}{\textbf{tensordot}}
\newcommand{\numssum}{\textbf{sum}}
\newcommand{\einsum}{\textbf{einsum}}
\newcommand{\axis}{\text{axis}}
\newcommand{\axes}{\text{axes}}

\newcommand{\g}{\textbf{g}}
\renewcommand{\v}{\textbf{v}}
\newcommand{\e}{\textbf{e}}
\newcommand{\x}{\textbf{x}}
\newcommand{\y}{\textbf{y}}

\renewcommand{\v}{\textbf{v}}

\newcommand{\M}{\textbf{M}}

\renewcommand{\S}{\textbf{S}}
\newcommand{\T}{\textbf{T}}
\newcommand{\B}{\textbf{B}}
\newcommand{\A}{\textbf{A}}
\newcommand{\C}{\textbf{C}}
\newcommand{\N}{\textbf{N}}
\newcommand{\X}{\textbf{X}}
\newcommand{\Y}{\textbf{Y}}
\newcommand{\Z}{\textbf{Z}}

\renewcommand{\H}{\textbf{H}}

\newcommand{\bop}{\bullet}

\DeclareMathSymbol{:}{\mathord}{operators}{"3A}

\newcommand{\project}{NumS}

\begin{document}

\title{\Large \bf NumS: Scalable Array Programming for the Cloud}


\author{Melih Elibol}
\author{Vinamra Benara}
\author{Samyu Yagati}
\author{Lianmin Zheng}
\author{Alvin Cheung}
\author{Michael I. Jordan}
\author{Ion Stoica}
\affil{\textit{University of California, Berkeley}}




\maketitle
\begin{abstract}
Scientists increasingly rely on Python tools to perform scalable distributed memory array operations using rich, NumPy-like expressions.
However, many of these tools rely on dynamic schedulers optimized for abstract task graphs, which often encounter memory and network bandwidth-related bottlenecks due to sub-optimal data and operator placement decisions. 
Tools built on the message passing interface (MPI), such as ScaLAPACK and SLATE, have better scaling properties, but these solutions require specialized knowledge to use.

In this work, we present \project, an array programming library which optimizes NumPy-like expressions on task-based distributed systems. This is achieved through a novel scheduler called Load Simulated Hierarchical Scheduling (LSHS). LSHS is a local search method which optimizes operator placement by minimizing maximum memory and network load on any given node within a distributed system. Coupled with a heuristic for load balanced data layouts, our approach is capable of attaining communication lower bounds on some common numerical operations, and our empirical study shows that LSHS enhances performance on Ray by decreasing network load by a factor of 2$\times$, requiring 4$\times$ less memory, and reducing execution time by 10$\times$ on the logistic regression problem.
On terabyte-scale data, \project\ achieves competitive performance to SLATE on DGEMM, up to 20$\times$ speedup over Dask on a key operation for tensor factorization, and a 2$\times$ speedup on logistic regression compared to Dask ML and Spark's MLlib.


\end{abstract}







\pagestyle{plain}
\section{Introduction}
\label{sec:intro}

\begin{figure}[ht]
    \centering
    \includegraphics[width=0.25\textwidth]{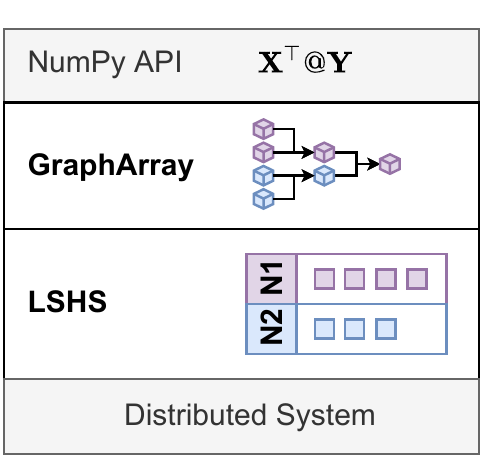}
    \caption{A high-level design diagram of \project. Users express numerical operations using the NumPy API. These operations are implemented by the GraphArray type. LSHS dispatches these operations to the underlying distributed system, specifying data and operator placement requirements. The underlying distributed system moves arrays between nodes and processes to satisfy data dependencies for task execution. Purple blocks reside on node $N_1$, and the blue blocks reside on node $N_2$.}
    \vspace{-0em}
    \label{fig:design}
\end{figure}

Scientists increasingly rely on modern Python tools ~\cite{xarray, dask} to perform scalable distributed numerical operations using rich, Numpy-like expressions.
However, many of these tools rely on dynamic schedulers~\cite{ray, dask} optimized for abstract task graphs, which do not exploit the structure of distributed numerical arrays, such as apriori knowledge of input and output sizes, or locality optimizations for parallel execution. This can lead to performance problems which are difficult to address without in-depth knowledge of the underlying distributed system.
These libraries implement array operations by constructing discrete tasks graphs which represent the desired computation, and schedule these tasks dynamically. While this decoupling of algorithm from scheduling has desirable software design properties, the loss of information to the scheduler leads to sub-optimal data and operator placement. In particular, when data placement is not optimized for numerical array operations, unnecessary communication among processes is often required in order to carry out basic operations, such as element-wise addition and vector dot products.

\begin{figure}[ht]
    \centering
    \includegraphics[width=0.25\textwidth]{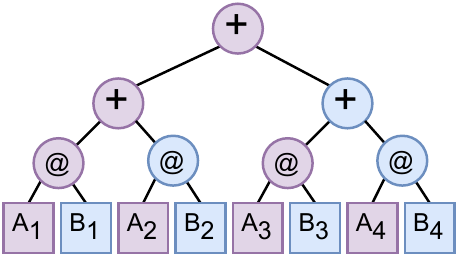}
    \vspace{-0em}
    \caption{A computation graph depicting potential sub-optimal placement of the data and operations required to perform $\A^\top @ \B$, where both $\A$ and $\B$ are partitioned row-wise into 4 blocks each. Transpose is fused with subsequent operations. In a 2-node cluster consisting of 4 workers each, a dynamic scheduler is unable to provide any locality guarantees on initial or intermediate data placement. Purple vertices correspond to data/operations placed on node 1, and blue vertices correspond to data/operations placed on node 2.}
    \vspace{-0em}
    \label{fig:inner-product-example}
\end{figure}

Figure \ref{fig:inner-product-example} illustrates a relatively simple example of what may occur when array operations are dynamically scheduled. A common operation which occurs when optimizing generalized linear models using Newton's method is $\A^\top @ \B$, where $@$ is the matrix multiplication operator in Python. 
When data and operator placement is delegated to a distributed system with dynamic scheduling, operands are often not co-located, requiring a greater amount of memory and data movement. Our example illustrates what happens in Dask for this operation (see Figure \ref{fig:micro-ablation} for empirical results). Independent tasks are executed round-robin over workers. Thus, all partitions of $\A$ are created on node 1, and all partitions of $\B$ are created on node 2. In the worst case, every operation requires at least one object transfer between nodes, requiring $3\times$ more time than the theoretical lower bound we provide within our framework. 

\begin{table}[ht]
\centering
\begin{tabular}{| r | l |}
\hline
\textbf{Syntax} & \textbf{Description} \\
\hline
 $-\X$ & Unary operation. \\ 
\hline
 $\X + \Y$ & Binary element-wise. \\ 
\hline
 $\numssum(\X, \axis=0)$ & Reduction. \\ 
\hline
 $\X @ \Y$ & Basic linear algebra. \\ 
\hline
 $\tensordot(\X, \Y, \axes=2)$ & Tensor contraction. \\
\hline
$\einsum(ik,kj->ij, \X, \Y)$ & Einstein summation. \\
\hline
\end{tabular}
\caption{\label{fig:operations} While \project\ provides greater coverage of the NumPy API, we list only the set of operations we consider in this work, along with syntax examples.}
\vspace{-0em}
\end{table}

Statically scheduled task-based distributed systems, such as Spark \cite{spark}, provide high-level APIs for machine learning. Spark's MLlib performs optimal scheduling for row-partitioned tall-skinny matrices, but does not provide support for block-partitioned multidimensional array programming.

High-performance computing (HPC) tools built on the message passing interface (MPI) ~\cite{mpi}, such as ScaLAPACK~\cite{scalapack} and SLATE~\cite{slate}, implement statically scheduled algorithms that are optimized for linear algebra operations. These specialized libraries are state-of-the-art for scalable linear algebra, but they do not support the NumPy API, making them inaccessible to an increasing number of scientists that use Python.

To enhance the performance of NumPy-based numerical array programming libraries, we present \project, a numerical array library optimized for task-based distributed systems. \project\ employs a scheduling framework tailored to the architecture and primitives provided by task-based distributed systems. Our framework optimizes data and operator placement decisions over a collection of compute nodes comprised of worker processes. As placement decisions are simulated or dispatched, our framework models memory load, network load, and object locality on each node and worker process. Within this framework, we implement a simple greedy operator placement algorithm guided by a cost function of our model of the underlying system's state, called Load Simulated Hierarchical Scheduling (LSHS) (see Section \ref{sec:lshs}). Our cost function computes the maximum memory and network load on any given node, which when minimized results in memory balanced and communication avoiding scheduling.

To support our empirical analysis, we analyze the communication time between node and worker processes, as well as the latency associated with dispatching placement decisions. We present communication lower bounds for a collection of common operations, and show that LSHS attains these bounds for some operations. We also show, both analytically and empirically, limitations of these distributed systems as they apply to the scalability of distributed numerical array operations.

Figure \ref{fig:design} depicts the design of \project. While \project\ operates on Dask, it is optimized for and achieves peak performance on Ray. When we refer to \project\ without explicitly stating the underlying distributed system, we assume it is using LSHS and running on Ray.

Overall, our evaluation shows that \project\ can achieve competitive performance to SLATE on square matrix multiplication on a terabyte of data, and outperforms Dask ML and Spark MLlib on a number of end-to-end applications. We show that, in every benchmark, LSHS is required for \project's algorithms to achieve peak performance on Ray.

In comparison to SLATE, \project\ is limited by the rate at which operations can be dispatched to the underlying distributed system, and the overhead introduced by remote function invocation.
Our empirical and theoretical analysis on these limitations show that
\project\ performs best on a smaller number of coarse-grained array partitions.
Fewer partitions mitigate control overhead introduced by having a centralized control process, and larger partitions amortize the overhead associated with executing remote functions in Ray.
Section \ref{sec:analysis} measures these overheads and provides a theoretical framework to model them in our analyses.
On the other hand, SLATE runs on MPI, which can operate on a greater number of partitions due to MPI's programming model, which enables decentralized and partitioned application control. 
For matrix multiplication, SLATE uses SUMMA, which allocates a memory buffer for output partitions, and accumulates intermediate results to achieve better memory efficiency than \project's matrix multiplication algorithm. 
However, SUMMA achieves sub-optimal communication time for a variety of matrix multiplication operations, such as matrix-multiplication among row-partitioned tall-skinny matrices. 
SLATE users must therefore be knowledgeable about the performance trade-offs of the variety of different operations provided by the library in order to take full advantage of its capabilities.
We therefore recommend \project\ for high-performance, medium-scale (terabytes) projects which may benefit from the flexibility of the NumPy API, and SLATE for projects which require specialized, large-scale distributed memory linear algebra operations.

Our contributions include:
\begin{enumerate}
    \item LSHS, a scheduling algorithm for numerical array operations optimized for task-based distributed systems.
    \item Theoretical lower bounds on the number of bytes required to perform a variety of common operations, including square matrix multiplication, as well as theoretical limitations of distributed numerical array computing on Ray and Dask.
    \item An ablation of LSHS, which shows that LSHS consistently enhances performance on Ray and Dask.
    On the logistic regression problem, LSHS enhances performance on Ray by decreasing network load by a factor of 2$\times$, using 4$\times$ less memory, and decreasing execution time by a factor of 10$\times$. 
    \item A comparison of \project\ to related solutions, showing that \project\ achieves a speedup of up to 2$\times$ on logistic regression compared to Dask ML and Spark's MLlib on a terabyte of data, and up to 20$\times$ speedup over Dask Arrays on core tensor factorization operations on a 4 terabyte tensor.
\end{enumerate}

\section{Related Work} \label{sec:related}
NumPy~\cite{numpy} provides a collection of serial operations, which include element-wise operations, summations, and sampling. When available, NumPy uses the system's BLAS~\cite{blas} implementation for vector and matrix operations. BLAS implementations use shared-memory parallelism.
For large datasets that fit on a single node, \project\ outperforms NumPy on creation and element-wise operations. NumPy does not provide a block partitioned representation of arrays on distributed memory.

Dask~\cite{dask} provides parallelism on a single machine via futures, and is able to scale to multiple nodes via the Dask distributed system, a distributed system framework similar to Spark~\cite{spark} and Ray~\cite{ray}.
Dask provides a distributed array abstraction, partitioning arrays along $n$-dimensions, and providing an API which constructs task graphs of array operations. When a graph of array operations is executed, array partitions and their operations are dynamically scheduled as tasks in an task graph. Round-robin scheduling of independent tasks results in sub-optimal data layouts for common array operations, such as the element-wise and linear algebra operations.
The Dask ML \cite{dask} library provides several machine learning models. The optimization algorithms written for these models frequently execute code on the driver process. The library is written using Dask's array abstraction.

Spark's MLlib~\cite{sparkmllib} is a library for scalable machine learning. MLlib depends on Breeze~\cite{breeze}, a Scala library that wraps optimized BLAS~\cite{blas} and LAPACK~\cite{lapack} implementations for numerical processing. 
Breeze provides high-quality implementations for many common machine learning algorithms that have good performance, but because it relies on Spark primitives, it introduces a learning curve for NumPy users.

Ray~\cite{ray} is a dynamically executing task-based distributed system. A Ray cluster is comprised of a head node and worker nodes. Worker nodes are comprised of worker processes, which execute tasks. The output of a task is written to the shared-memory object store on the node on which the task was executed. Any worker can access the output of any other worker within the same node. Python programs connect to a Ray cluster via a centralized driver process, which dispatches remote function calls to a local scheduler which schedules operations on worker processes.
Ray implements a \textit{bottom-up} distributed scheduler. Driver and worker processes submit tasks to their local scheduler. In general, based on available resources and task meta data, a local scheduler may execute a task locally, or forward the task to a centralized process on the head node. The latter occurs only when a local scheduler is unable to execute a task locally, a decision based on a variety of heuristics. In general, when a local scheduler is presented with a collection of tasks which have no dependencies, it distributes tasks to reduce overall load on any given node.


High Performance Computing (HPC) libraries such as ScaLAPACK (Scalable Linear Algebra PACKage) ~\cite{scalapack} and SLATE~\cite{slate} provide tools for distributed memory linear algebra. They implement highly optimized communication-avoiding operations using MPI~\cite{mpi}. These libraries are state-of-the-art in terms of performance. The high performance provided by these libraries comes at the cost of several specialized implementations of various linear algebra operations. ScaLAPACK exposes 14 different routines for distributed matrix multiplication, each optimized for specialized matrices with various properties~\cite{scalapack}. Unlike \project\ which enables programming against the NumPy API, which includes support for tensor algebra operations, while these libraries provide a C++ API limited to linear algebra operations.



Deep learning libraries such as Tensorflow~\cite{abadi2016tensorflow}, PyTorch~\cite{paszke2019pytorch}, and MXNet~\cite{chen2015mxnet} provide tensor abstractions, and JAX~\cite{jax} provides a NumPy array abstraction which enhances usability.
Mesh Tensorflow \cite{meshtf} provides tensor partitioning on top of Tensorflow, but requires specifying layouts for tensors, and targets Google TPUs.
These libraries are specialized for DNN training on accelerators. In contrast, \project\ is designed for general purpose array programming on CPUs.



\section{Background}
\label{sec:background}

In task-based distributed systems, a \textit{task} is a unit of work with one or more dependencies. This can be thought of as an arbitrary pure function.
\textit{Objects} serve as the inputs and outputs of tasks. A completed task can be viewed as the object(s) it outputs.
A \textit{task graph} is a representation of the dependencies between tasks.
A \textit{remote function call} (RFC) creates a task with zero or more dependencies. In \project, numerical kernel operations are executed as RFCs, creating a task with the kernel operation's operands as dependencies. Operands may be tasks that have not yet been executed, or they may be the output object of a completed task.

A \textit{worker} is an independent process which executes tasks. We use the term \textit{node} to refer to machines comprised of one or more worker processes.
\textit{Task placement} refers to the process of deciding on which node or worker a particular task should execute. All task placement decisions are executed on a centralized \textit{driver process}. 

\begin{wrapfigure}{r}{0.25\columnwidth}
    \vspace{-1em}
    \includegraphics[width=0.25\columnwidth]{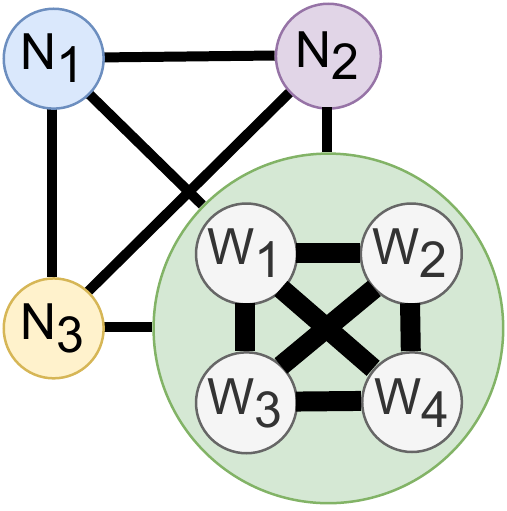}
    \caption{}
    \vspace{-1em}
    \label{fig:network-topology}
\end{wrapfigure}
Figure \ref{fig:network-topology} depicts the network topology for which \project\ is optimized. The cluster is comprised of 4 nodes, with 4 workers per node. Node $N_3$ is expanded to expose the intra-node network topology of worker processes. Thicker edges correspond to greater bandwidth. 
In Ray, we make placement decisions at the granularity of nodes, leaving worker-level scheduling to each node's local scheduler. Ray implements a shared memory object store, enabling any local worker to access the output of any other local worker without worker-to-worker communication. In Dask, we make placement decisions at the granularity of worker processes. worker-to-worker communication within the same node can be expensive, which we address with our hierarchical data and operator placement design. Our design is given in Section \ref{sec:grapharray}.

\section{Graph Arrays}
\label{sec:grapharray}

The \textbf{GraphArray} type implements distributed array creation, manipulation, and numerical operations. Creation and manipulation operations \textit{execute immediately}, whereas numerical operations are deferred.

A GraphArray is created via read operations, invocation of operations like \textbf{zeros}(shape, grid) to create a dense array of zeros or ones, or operations like \textbf{random}(shape, grid) to randomly sample a dense array from some distribution. The \textit{shape} parameter specifies the dimensions of the array, and the \textit{grid} parameter specifies the \textit{logical partitioning} of the multidimensional array along each axis specified by the shape. The logical partitioning of an array is called its \textit{array grid}. For example, $\A = \textbf{random}((256, 256), (4, 4))$ will randomly sample a block-partitioned array partitioned into 4 block along the first axis, 4 blocks along the second axis, and 2 blocks along the third axis for a total of 16 blocks. 
We use the notation $\A_{i,j}$ to denote the $i,j$ block of $\A$, as depicted in Equation \ref{eq:blocks}. Each block $\A_{i,j}$ {\it is itself a matrix with dimensions $64 \times 64$}.
\begin{align}
\A & =
\begin{bmatrix}
    \A_{0, 0} & \A_{0, 1} & \A_{0, 2} & \A_{0, 3}  \\
    \A_{1, 0} & \A_{1, 1} & \A_{1, 2} & \A_{1, 3}  \\
    \A_{2, 0} & \A_{2, 1} & \A_{2, 2} & \A_{2, 3}  \\
     \A_{3, 0} & \A_{3, 1} & \A_{3, 2} & \A_{3, 3}
 \end{bmatrix}
\label{eq:blocks}
\end{align}

When a grid argument is not specified, \project\ chooses a partitioning according to the \textit{softmax distribution of the array's dimensions}. for a vector $\x \in \mathbb{R}^n$, the softmax function is $\sigma(\x)_i = \frac{\e^{\x_i}}{\sum_{j=0}^{n-1}{\e^{\x_j}}}$. We use this distribution to factor the total number of available worker processes into the number of dimensions of an array, putting greater weight on larger dimensions. If we have $p=16$ worker processes, we automatically set the grid to $p^{\sigma(shape)} = (4, 4, 1)$. This approach partitions tall-skinny matrices along its larger axis, and provides balanced partitioning for square or near-square matrices.
\begin{figure}[ht]
    \centering
    \begin{subfigure}[b]{0.40\columnwidth}
      \centering
      \includegraphics[width=0.9\textwidth]{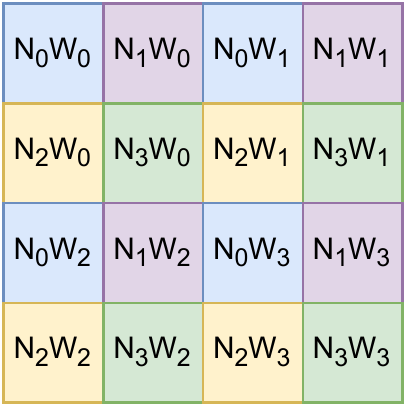}
      \caption{Worker Mapping.}
        \vspace{-0em}
      \label{fig:worker-grid}
    \end{subfigure}
    ~
    \begin{subfigure}[b]{0.40\columnwidth}
      \centering
      \includegraphics[width=0.9\textwidth]{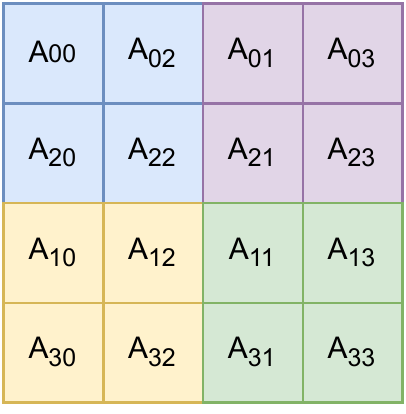}
      \caption{Array Mapping.}
      \vspace{-0em}
      \label{fig:array-mapping}
    \end{subfigure}
    \caption{Hierarchical mapping of logical array partitions to physical nodes and workers.
    Within a $(2, 2)$ node grid, blue, purple, yellow, and green correspond to nodes $(0,0), (0,1), (1,0), (1,1)$, respectively.
    }
    \vspace{-0em}
    \label{fig:hierarchical-scheduling}
\end{figure}

When a creation operation is invoked, the logical partitions of an array are mapped \textit{hierarchically} to physical nodes. To carry out this mapping within our framework, a user-defined \textit{node grid}, a multi-dimensional coordinate space for nodes within a cluster, is required.
For a cluster consisting of $4$ nodes with $4$ workers each, figure \ref{fig:worker-grid} depicts the mapping of the previously defined array $\A$ to nodes and workers from a user-defined $(2, 2)$ (sometimes written $2 \times 2$) node grid.
$N_{i}W_j$ corresponds to worker $j$ on node $i$. The node grid is fixed throughout the execution of a \project\ application.
For a node grid with dimensions $g_1 \times g_2$, $\A_{i,j}$ is placed on node $N_\ell$, where $\ell=(i\%g_1) g_2 + j\%g_2$. Within each node, blocks of $\A$ are placed round-robin over available workers. In Ray, worker-level mapping is ignored since worker processes on the same node use a shared-memory object store.
In our example, $\A_{2, 3}$ is placed on node $N_{1}$ and worker $W_{3}$: The node placement is straightforward, whereas the worker-level placement is due to 3 other partitions which are placed on node $N_1$. Figure \ref{fig:array-mapping} depicts the grouping of partitions within each node. 
We call this approach to cyclic mapping of partitions, both over nodes and workers, a \textit{hierarchical data layout}.
Along a particular dimension, when operands have the same shape and grid parameters, our data layout co-locates operand partitions, requiring zero communication for element-wise operations. This layout also minimizes communication for a variety of linear and tensor algebra operations. With this data layout, the operation in Figure \ref{fig:inner-product-example} requires zero communication to achieve the first set of matrix multiplication operations (minimizing inter-node communication for sums is presented in Section \ref{sec:lshs}).


%



\begin{figure}[h]
    \vspace{-1.5em}
    \centering
    \begin{subfigure}[b]{0.31\columnwidth}
      \centering
      \includegraphics[width=0.17\textwidth]{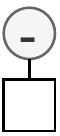}
      \caption{$-\X$}
      \label{fig:ops-neg}
    \end{subfigure}
    ~
    \begin{subfigure}[b]{0.31\columnwidth}
      \centering
      \includegraphics[width=0.35\textwidth]{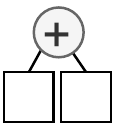}
      \caption{$\X+\Y$}
      \label{fig:ops-add}
    \end{subfigure}
    ~
    \begin{subfigure}[b]{0.31\columnwidth}
      \centering
      \includegraphics[width=0.75\textwidth]{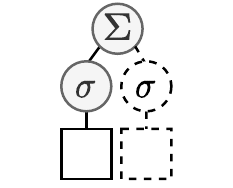}
      \caption{$\numssum$}
      \label{fig:ops-sum}
    \end{subfigure}
    ~\\
    \begin{subfigure}[b]{0.31\columnwidth}
      \centering
      \includegraphics[width=0.6\textwidth]{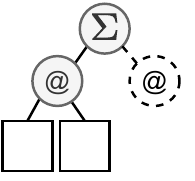}
      \caption{$\X@\Y$}
      \label{fig:ops-matmul}
    \end{subfigure}
    ~
    \begin{subfigure}[b]{0.31\columnwidth}
      \centering
      \includegraphics[width=0.6\textwidth]{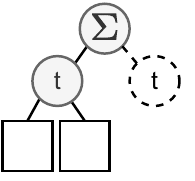}
      \caption{$\tensordot$}
      \label{fig:ops-tensordot}
    \end{subfigure}
    ~
    \begin{subfigure}[b]{0.31\columnwidth}
      \centering
      \includegraphics[width=0.80\textwidth]{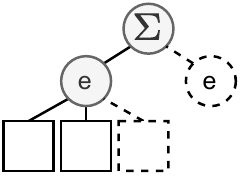}
      \caption{$\einsum$}
      \label{fig:ops-einsum}
    \end{subfigure}

    \caption{
    Subgraphs induced by operations listed in \ref{fig:operations} within a GraphArray. Rectangular vertices correspond to leaf vertices, and circular vertices correspond to operations. Dashed lines are used to denote repetition. 
    $\Sigma$ corresponds to $Reduce(add, \dots)$, $\sigma$ corresponds to $ReduceAxis(add, \X, axis)$, and $@$, $t$, $e$ correspond to matrix multiplication, $\tensordot$, and $\einsum$, respectively.}
    \vspace{-0em}
    \label{fig:ops}
\end{figure}

While creation operations are immediately executed, arithmetic operations in \project\ are \textit{lazily} executed. Figure \ref{fig:ops} provides the subgraphs induced when a particular operation is performed among GraphArray's consisting of leaf vertices. When an operation is performed on one or more GraphArrays, an array of subgraphs is generated, depicting the sub-operations required to compute the operation. Operations performed on GraphArrays are referred to as \textit{array-level} operations, whereas the sub-operations performed among leafs and vertices are referred to as \textit{block-level} operations. The operands involved in \textit{block-level} operations are referred to as \textit{blocks}. Blocks are either materialized subarrays, or \textit{future} subarrays which have not yet been computed.

To perform the $-\X$, a new GraphArray is constructed, and each leaf vertex is replaced by the subgraph given in Subfigure \ref{fig:ops-neg}. 
For $\X + \Y$, the dimensions of $\X$ and $\Y$ and their grid decompositions are required to be equivalent. The execution of this expression generates a new GraphArray, comprised of an array of subgraphs given by Subfigure \ref{fig:ops-add}.

The $\numssum(\X, axis)$ operation depends on the $Reduce(add, \dots)$ and $ReduceAxis(add, \X, axis)$ vertex types. The $ReduceAxis$ vertex type takes a single block and reduces it using the given operation and axis. The $Reduce$ vertex type takes any number of blocks with equivalent dimension, and reduces them using a given operation. The $\numssum$ operation is implemented by first summing each individual block along the given axis, and then summing output blocks along the same axis. For example, if we performed $\A' = \numssum(\A, 0)$ on the array we previously defined, the output blocks of $\A'$ would be vectors with dimension $64$, and its array grid would be single-dimensional, consisting of $4$ blocks.

\begin{figure}[ht]
    \centering
  \centering
  \includegraphics[width=0.95\columnwidth]{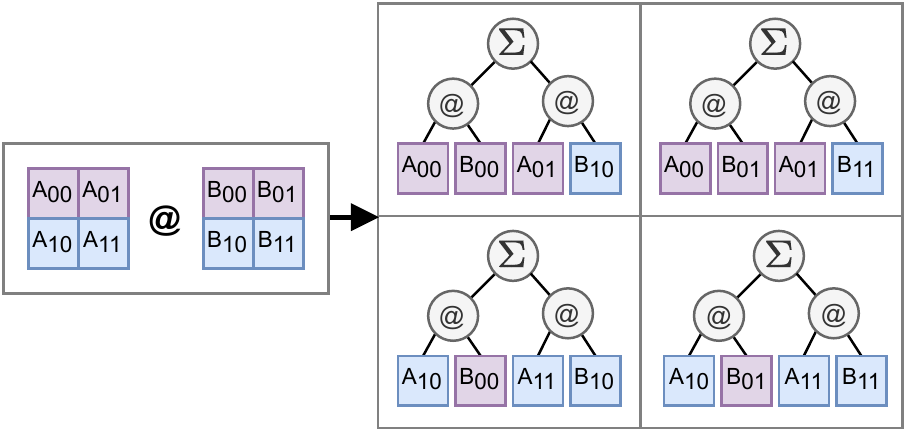}
  \caption{Matrix multiplication $\A @ \B$ of two arrays $\A$ and $\B$ partitioned into $2\times2$ array grids. The operation is invoked on a cluster with node grid $2 \times 1$. Blocks on node $(0,0)$ are colored purple, and blocks on node $1,0$ are colored blue.}
    \vspace{-0em}
    \label{fig:matmul-example}
    \vspace{-0em}
\end{figure}

Figure \ref{fig:matmul-example} provides a simple example of matrix multiplication. Matrix multiply is broken down into independent sub-matrix multiplication operations. The output of each sub operation is summed used the $Reduce(add, \dots)$ vertex type.

The rest of the operations given in Figure \ref{fig:ops} are structurally similar to $\numssum$ and matrix multiplication. The operations are broken down into sub-operations of the same kind, and a $Reduce(add, \dots)$ vertex is used to sum the intermediate outputs. In this sense, these operations can be viewed as \textit{recursive}.

When two or more sub-operations generated by these induced subgraphs require operands which are located on different nodes, the operation may be executed, or \textit{placed}, on multiple nodes. The decision process for placing operations is left to our scheduling algorithm, which is described in Section \ref{sec:lshs}, but LSHS requires a set of \textit{placement options} to be provided by every vertex in a GraphArray. For unary, and $ReduceAxis$ operations, there is only a single operand. For element-wise binary operations, the data is already located on the same node and workers by our hierarchical mapping procedure, so only one potential option is provided to the scheduler for these vertex types. For matrix multiplication, tensor dot, and einsum, the set of placement options is the union of all the nodes on which all the operands reside.

The $Reduce$ vertex may have $n$ operands. Our scheduler must place $n-1$ binary operations in order to complete the execution of the $Reduce$ vertex. For each of the $n-1$ binary operations, the set of placement options is the union of all the nodes on which each of the two operands reside. The $Reduce$ vertex is responsible for deciding which operands to pair for each of the $n-1$ binary operations. We pair operands according to their locality within the hierarchical network. We first pair operands on the same workers, then we pair operands on the same node. Our scheduler must make placement decisions for operands which are not on the same workers.
We describe our solution to this scheduling problem in the next section.










\section{Load Simulated Hierarchical Scheduling}
\label{sec:lshs}
\begin{figure*}
    \centering
    \includegraphics[width=.95\textwidth]{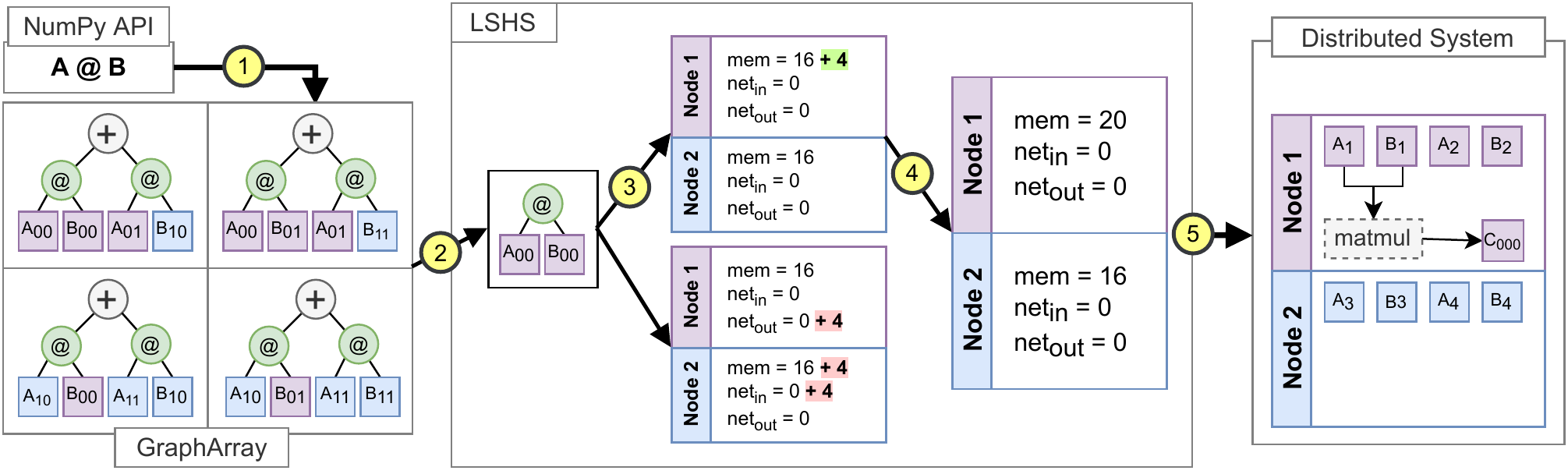}
    \vspace{-0em}
    \caption{Scheduling of the expression $\A @ \B$ on a $2$ node cluster, where $\A$ and $\B$ are both $4\times4$, and both have a block shape of $2\times2$. 1) The expression is executed at runtime. 2) A frontier vertex is sampled at random. 3) Operator placement is simulated on each node. 4) The operation is placed on the node which minimizes the objective. 5) The distributed system executes the operation on the given node.
    }
    \vspace{-0em}
    \label{fig:nums-execution}
\end{figure*}
\begin{algorithm}[h]
\SetKwFunction{lshs}{lshs}
\SetKwFunction{frontier}{frontier}
\SetKwFunction{size}{size}
\SetKwFunction{sample}{sample}
\SetKwFunction{nodes}{nodes}
\SetKwFunction{transition}{transition}
\SetKwFunction{cost}{cost}
\SetKwProg{Pn}{Function}{:}{\KwRet}
\Pn{\lshs{s}}{
    \While{$\frontier(s)$}{
        $N_{min} \gets null$ \;
        $C_{min} \gets \infty$ \;
        $v = \sample(\frontier(s))$ \;
        \For{$i \gets 0$ \KwTo $k$}{
            \uIf{$\cost{v, N_i} < C$}{
                $C_{min} = \cost{s, N_i}$ \;
                $N_{min} = N_i$ \;
            }
        }
        $s \gets \transition(s, N_{min})$ \;
    }
    \KwRet s \;
}
\caption{LSHS.}
\label{algo:lshs}
\end{algorithm}
LSHS approximately minimizes an optimization-based formulation of operator scheduling within a distributed system. We describe the procedure for Ray, defining the placement procedure over nodes.
There are three primary components to LSHS: A GraphArray; a cluster state object used to simulate load imposed on the cluster for a particular placement decision; and an objective function which operates on an instance of the cluster state. LSHS is a discrete local tree search algorithm  (see Algorithm \ref{algo:lshs})~\cite{randn}. LSHS executes the GraphArray $s$ by sequentially scheduling \textit{frontier} vertices. An operation vertex is on the frontier when all of its children are leaf vertices. A vertex is sampled from the frontier, and the placement option which minimizes the cost function is selected. The GraphArray is then \textit{transitioned} to a new GraphArray by either updating a $Reduce$ vertex to reflect its remaining child operands, or converting an operation vertex to a leaf vertex. 
The algorithm terminates when $s$ consists of all leaf vertices. For each output graph, the last operation is mapped according to the hierarchical data layout, which ensures that every graph array has a hierarchical data layout. This is implicitly handled within the transition function.

Figure ~\ref{fig:nums-execution} provides a step-by-step depiction of the execution of the matrix multiplication operation given in \ref{fig:matrix-multiply}. Each step in the algorithm is depicted by a numerically labeled arrow. 
Step 1 generates the GraphArray as described in Section \ref{sec:grapharray}. The vertices which are highlighted green are frontier nodes. Step 2 randomly samples a frontier node. In step 3, each placement option is simulated, and its costs is computed. In step 4, the option that minimizes the cost function is chosen, and the GraphArray is transitioned to its new state. In Step 5, we show how the transition procedure performs the actual remote function call to the underlying distributed system, which carries out the required operation on Node 1. Eventually, for each output graph, the final operation required to complete the computation of $\A @ \B$ will be the addition operation. In this example, the addition operation is scheduled using the hierarchical data layout.

\subsection{Cluster State and Optimization Problem}
\label{subsubsec:crts-objective}

The cluster state, which is depicted in step 4 of Figure \ref{fig:nums-execution}, is used to monitor the memory and network load imposed on all nodes within a Ray cluster. To simplify exposition, we use the number of elements in an array to signify both memory and network load \footnote{An additional coefficient we use to discount worker-to-worker communication within the same node on Dask. Ray does not require such a coefficient since workers operate on a shared-memory object store within each node.}. For a given node, we compute the network load as two integers: The total number of incoming and outgoing array elements. The memory load is the total number of array elements on a node resulting from the transmission of arrays to that node, and the output of any operation executed on that node. Let $\M$ denote a data structure that maintains a mapping from all objects to their corresponding nodes, and $\S$ denote a $k\times3$ matrix maintaining the memory, network in, and network out of a $k$ node cluster. Let $m=0, i=1, o=2$ so that $\S_{j, m}$ corresponds to the memory load on node $j$, and likewise $i$ corresponds to input load, and $o$ corresponds to output load.
Let $\A$ correspond to the set of scheduling actions (i.e. nodes on which to schedule an operation) available for vertex $\v$. An action $\a \in \A$ is a tuple $(j, size)$, where $j$ corresponds to the $j$th node in $\S$, and $size$ corresponds to the size of the output of vertex $\v$. Let $\S', \M' = \T(\S, \M, \a)$ be a transition function that takes $\M, \S, \a$ and returns $\S', \M'$ such that the operation $\v$ is simulated on $\S$ via the action $\a$. With $\M$, the transition function $\T$ has enough information to simulate object transfers between nodes.

At a given cluster state $\S, \M$, the objective function which obtains the best action $\a$ from the set of actions $\A$ available to vertex $\v$ is formulated as follows.
\begin{equation}
    \begin{aligned}
    \min_{\a \in \A} \,\, & \left (
    \max_{j=1}^{k} {\S'_{j, m}} +
    \max_{j=1}^{k} {\S'_{j, i}} +
    \max_{j=1}^{k} {\S'_{j, o}}
    \right ) \\ 
    \text{subject to} & \,\,\,\, \S', \M' = \T(\S, \M, \a).
    \end{aligned}
    \label{eq:objective-function}
\end{equation}


If we modify Equation \ref{eq:objective-function} to jointly minimize the maximum memory and network loads over all nodes for a collection of operations, the problem is very similar to load balancing: Given a collection of task execution times, load balancing minimizes maximum execution time over a collection of nodes. 

To construct a reduction from load balancing, we need an optimization problem that minimizes the maximum load over all possible scheduling choices. We introduce the superscript $t$ to identify the state of the variables at a given step in the sequence of actions required to compute the optimal solution. The superscript $t$ is used to identify the vertex $\v^t$ being computed at step $t$. In addition to containing the node on which to compute vertex $\v^t$ and its output size, the set of actions $\A^t$ are expanded to include the next vertex $\v^{t+1}$ on which to operate. We incorporate these new requirements into the transition function $\T$ by returning the set of next actions $\A^{t+1}$ as well.
We represent the optimal solution as a sequence of actions $\pi$, where $\pi^t \in \A^t$. For a computation tree consisting of $n$ operations, the optimal sequence of actions of length $n$ is given as follows.
\begin{equation}
    \begin{aligned}
    \min_{\pi \in (\A^{0}, \,\dots\,, \,\, \A^{n-1})} & \left (\max_{j=1}^{k} {\S^{n}_{j, m}} + \max_{j=1}^{k} {\S^{n}_{j, i}} + \max_{j=1}^{k} {\S^{n}_{j, o}} \right ) \\
    \text{subject to} & \,\,\,\,\, \S^{t+1},\, \M^{t+1},\, \A^{t+1} = \T(\S^{t}, \M^{t}, \pi^{t}).
    \end{aligned}
    \label{eq:general-opt-problem}
\end{equation}
The optimization problem given in Equation \ref{eq:general-opt-problem} is NP-hard by a straightforward reduction from load balancing. Tasks in load balancing are independent. In our formulation, independent tasks require no object transfers between nodes. Thus, a load balancing problem instance can be converted to an instance of the problem given by Equation \ref{eq:general-opt-problem}, and all solutions to these problem instances will have zero network load. Instead of maximum memory load, the remaining term $\max_{j=1}^{k} {\S^{n}_{j, m}}$ is used to compute the maximum time for all tasks to execute on any given node.

\section{Generalized Linear Models}
\label{sec:glm}

\begin{algorithm}[h]
\SetAlgoLined
    $\beta \gets {\bf 0}$;\\
    \While{true}{
        $\mu \gets m(\X, \beta)$; \\
        $\g \gets \nabla f(\X, \y, \mu, \beta)$; \\
        $\H \gets \nabla^2 f(\X, \y, \mu, \beta)$ \\
        $\beta \gets \beta - \H^{-1} \g$; \\
        \If{$\| \g \|_2 \leq \epsilon$}{
            return $\beta$;
        }
    }
\caption{Newton's method~\cite{wright}.}
\label{algo:newton}
\end{algorithm}

Generalized linear models (GLMs) are notoriously difficult to scale due to their reliance on basic array and linear algebra operations. Furthermore, optimizing these models with second order methods require the expensive computation of the Hessian matrix, or approximations to the Hessian matrix \cite{wright, bishop}. Theoretically, Newton's method converges faster than any other method available for GLMs. In practice, fast convergence can be difficult to achieve without proper utilization of all computational resources when fitting these models to large amounts of data. \project\ is able to achieve high performance on any model which relies heavily on element-wise and basic linear algebra operations, making GLMs an ideal \project\ application.

Given a tall-skinny dataset $\X \in \mathbb{R}^{n \times d}$ decomposed into a grid $q \times 1$ of blocks, $\y \in \mathbb{R}^{n \times 1}$ decomposed into a grid $q \times 1$, a GLM $m$ with corresponding twice differentiable convex objective $f$, and minimum gradient norm $\epsilon$, Algorithm \ref{algo:newton} computes the global minimum $\beta \in \mathbb{R}^{d}$ of $f$ using Newton's method. Upon initialization, $\beta$ is decomposed into a $1 \times 1$ grid. 

We work through the execution of Algorithm \ref{algo:newton} on an $r \times 1$ grid of nodes. We use the notation $\N_{i,0}$ to refer to the $i$th node in the $r \times 1$ grid. We assume the hierarchical data layout, so that $\X$ and $\y$ are distributed row-wise over $r$ nodes, and the single block $\beta_{0, 0}$ of $\beta$ is created on node $\N_{0,0}$. In this example, we assume that expressions are computed upon assignment. For example, LSHS is applied to the computation tree induced by the expression $\mu = m(\X, \beta)$, which schedules the operations which comprise the computation tree and places the output $\mu$ using the hierarchical data layout. Similarly, the expressions assigned to $\g$, $\H$, and $\beta$ are represented as computation trees which are submitted to LSHS for scheduling.

For logistic regression, we have that $m(\X, \beta) = \frac{1}{1 + e^{-\X \beta}}$. We see first that the linear operation $\X \beta$ will yield an intermediate value $\C$ comprised of a grid of blocks $\C_{i, 0} = \X_{i, 0} \times \beta_{0, 0}$. It is usually the case that $\X_{i,0}$ has more elements than $\beta$, so LSHS will broadcast $\beta_{0, 0}$ to all the nodes on which the $\X_{i, 0}$ reside. LSHS will schedule the remaining unary and binary operations to the blocks of $\C$ in-place: Since all data is local, the node placement options are reduced to the node on which all of the data already resides. The output $\mu$ is decomposed into a $g_1 \times 1$ matrix and distributed according to the hierarchical data layout.

The gradient of $f$ is given by the expression $\X^\top (\mu - \y)$. We see that $\mu$ has the required grid decomposition to perform the element-wise subtraction operation with $\y$, yielding an intermediate vector ${\bf c}$. Because $\mu$ and $\y$ have the same size, partitioning, and layout, the data is local, and LSHS schedules the operations without data movement.
In \project, transpose is executed lazily by fusing with the next operation. Thus, the transpose operator is fused with the matrix-vector multiply, resulting in the final set of operations $\g_{0, 0} = \sum_{h=0}^{g_1-1} (\X^\top)_{0, h} {\bf c}_{h, 0}$. 
Like the element-wise operations, because blocks of $\X$ and ${\bf c}$ have the same size and partitioning along axis $0$, the product operation executes locally.
The sum of the products is executed as a reduction: A reduction tree is formed, and any local sums are executed first. The remaining sums are scheduled according to the cost function defined for LSHS. Like $\beta$, $\g$ is comprised of a single block, therefore the final sum is scheduled on node $\N_{0,0}$ to satisfy the hierarchical data layout required by the outputs of LSHS. 

The Hessian of $f$ is given by the expression $\X^\top (\mu \odot (1 - \mu) \odot \X)$, where $\odot$ denotes element-wise multiplication. The expression $\mu \odot (1 - \mu)$ is scheduled locally by LSHS because the input data for these sub trees all reside on the same node. The intermediate vector $\textbf{c}$ resulting from the previous operation has the same block decomposition as $\mu$, and $\mu$ has the same size and partitioning as the first dimension of $\X$, therefore the blocks of $\textbf{c}$ and $\X$ are distributed the same way over the $r$ nodes. Thus, the vector-matrix element-wise operation $\textbf{c} \odot \X$ is executed without data movement: NumPy executes this kind of expression by multiplying $\textbf{c}$ with every column of $\X$, yielding an intermediate matrix $\C$ with the same decomposition as $\X$.
Finally, the operation $\X^\top \C$ results in the computation $\H_{0,0} = \sum_{h=0}^{k-1} (\X^\top)_{0, h} \C_{h, 0}$, where $\H$ is square with dimension $d$. The matrix multiplications are executed without data movement, and the sum operation is executed as a reduction like the previous reduction. Since $\H$ is square with dimension $d$, it is single partitioned, and the final operation to compute $\H$ is scheduled to occur on node $\N_{0,0}$.

Finally, we update $\beta$. At this point, $\beta$, $\g$ and $\H$ are all comprised of single blocks, and all reside on node $\N_{0,0}$, so the update to beta is executed locally on node $\N_{0,0}$. Similar to the update of $\beta$, the gradient norm is computed locally on node $\N_{0,0}$.




\section{Communication Analysis}
\label{sec:analysis}

We extend the $\alpha-\beta$ model of communication to analyze the communication time of element-wise, reduction, and basic linear and tensor algebra operations. In our model, for a particular channel of communication, $\alpha$ denotes latency and $\beta$ denotes the inverse bandwidth of the channel. We also model the time to dispatch an operation from the driver process as $\gamma$. The time to transmit $n$ bytes between two nodes is given by $C(n) = \alpha + \beta n$. 
We also model the implicit cost of communication between workers within a single node of Ray as $R(n) = \alpha' + \beta' n$.
For a dense array of size $N$, let $p$ the number of workers, $N/p = n$ the block size, or number of elements, and $r = p/k$ the number of workers-per-node on a $k$ node cluster.


\begin{figure}[ht]
    \centering
    \begin{subfigure}[b]{0.45\columnwidth}
      \centering
      \includegraphics[width=\textwidth, trim=0cm 0cm 0cm 1.2cm, clip]{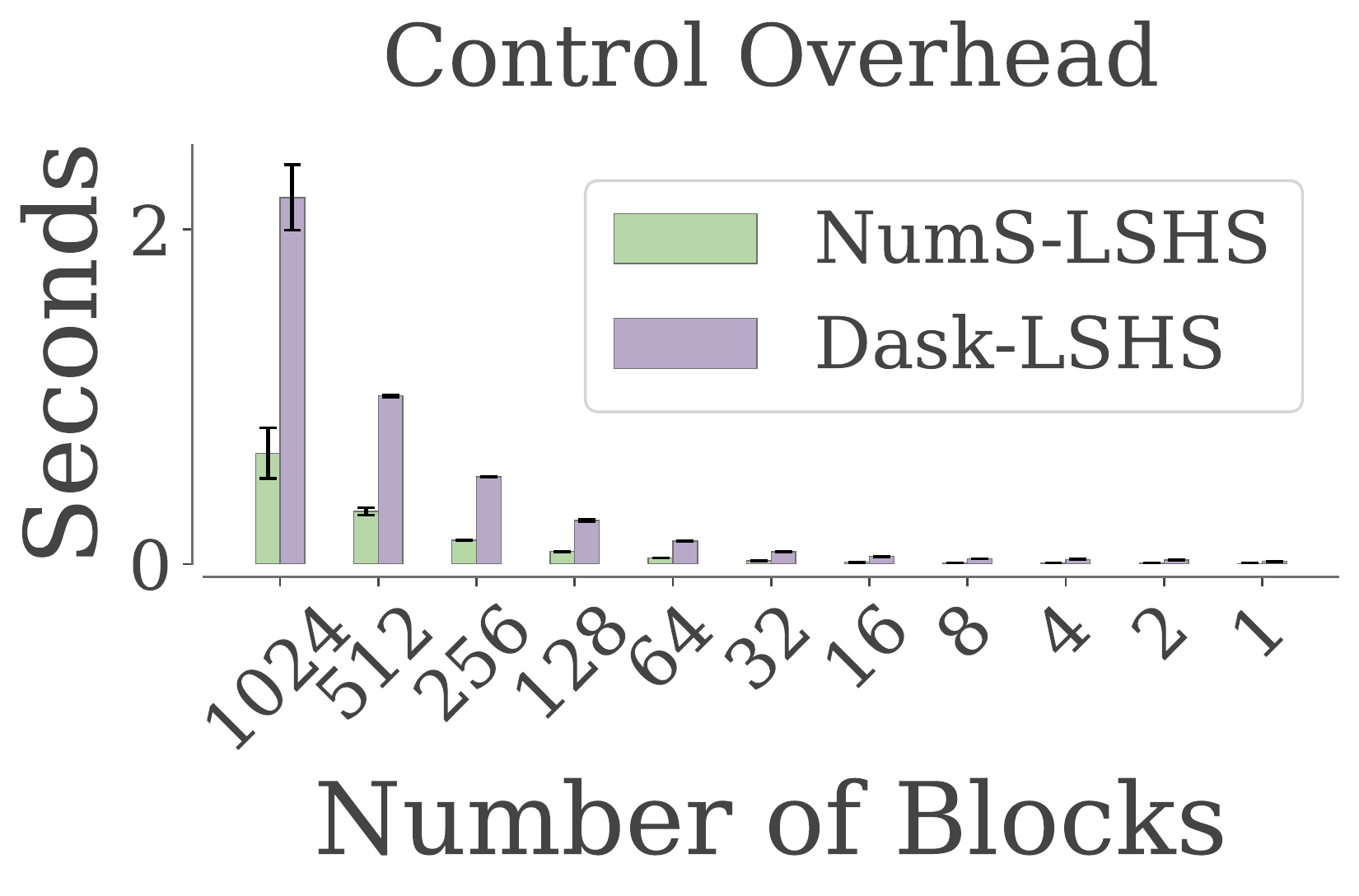}
      \caption{Control Overhead.}
      \label{fig:driver-limit}
    \end{subfigure}
    ~
    \begin{subfigure}[b]{0.45\columnwidth}
      \centering
      \includegraphics[width=\textwidth, trim=0cm 0cm 0cm 1.2cm, clip]{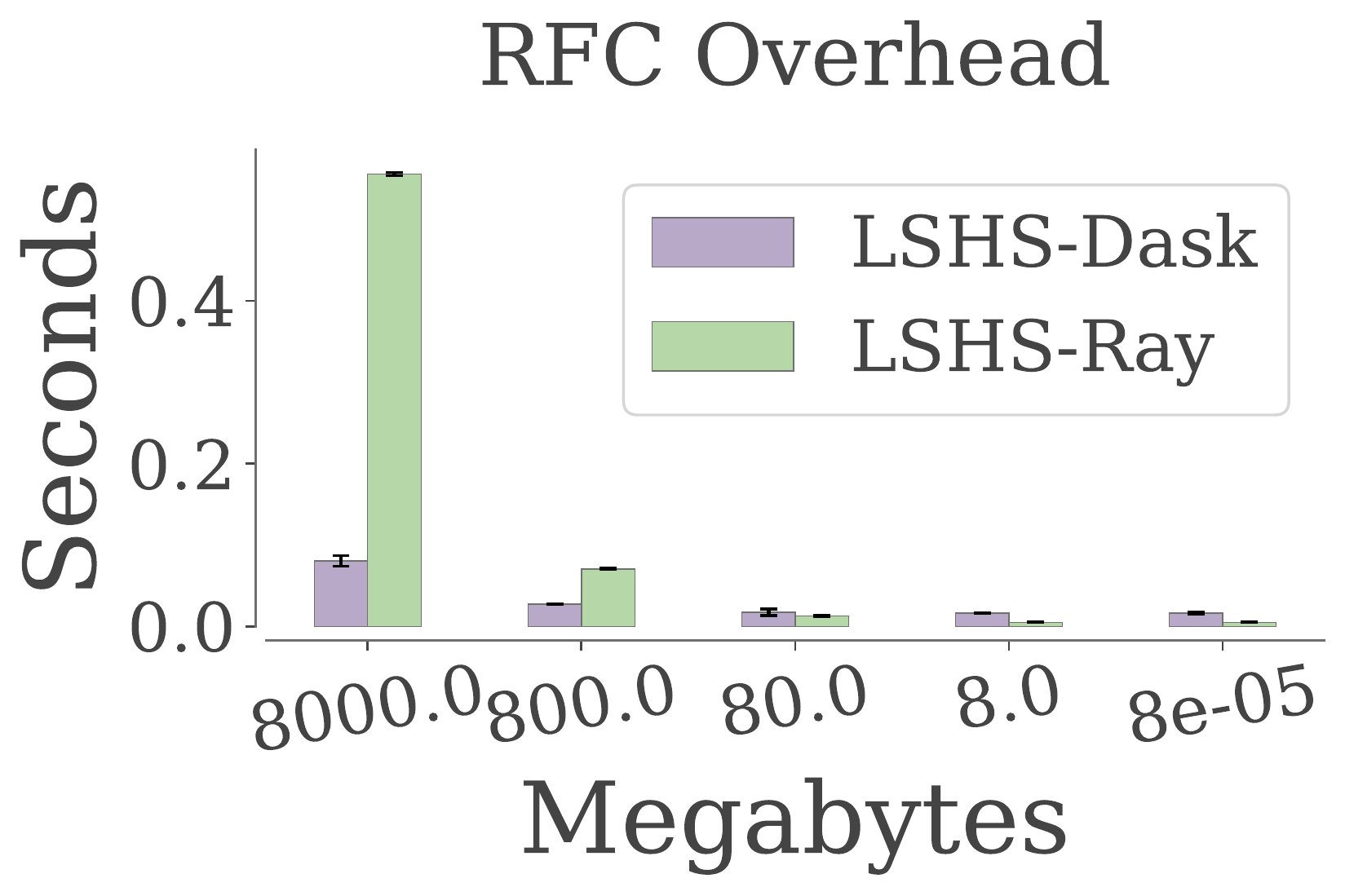}
      \caption{RFC Overhead.}
      \label{fig:rfc-limit}
    \end{subfigure}
    \vspace{-0em}
    \caption{
    Control overhead is measured by the time it takes to allocate a vector of dimension $1024$ on a 16 node cluster, with a total of $1024$ workers and cores. This is captured by the $\gamma$ term. As we decrease the number of blocks, $\gamma$ decreases.
    RFC overhead is measured by executing $-\x$ on a single block vector $\x$. The overhead is directly measured as the difference between the time it takes to perform this operation using NumPy. Ray writes task outputs to an object store, resulting in greater RFC overhead. This is captured by the $R(n)$ term.}
    \vspace{-0em}
    \label{fig:limits}
\end{figure}


The results of our theoretical analysis show that LSHS attains the lower bound $0$ for unary and binary element-wise operations. For row-partitioned $\X$ and $\Y$, we also attain the lower bound for $\numssum$ and $\X^\top \Y$, both of which are logarithmic in $k$, as well as $\X \Y^\top$. A complete analysis is given in Appendix \ref{appendix:analysis}. 

We are not able to provide an upper bound for square matrix multiplication for LSHS, but we do provide a lower bound which shows that LSHS is capable of achieving asymptotically faster communication time in $k$ (the number of nodes) than SUMMA.



\section{Evaluation}
\label{sec:eval}
\label{sec:setup}


LSHS enhances \project's performance by balancing data placement within multi-node cloud-based distributed systems, and placing operations in a fashion which maintains load balance while minimizing inter-node communication.
To evaluate the contribution of LSHS to \project's performance,
we evaluate \project\ with and without LSHS in a variety of benchmarks.
Partitioning plays a key role in performance. To begin with, we measure 
how partitioning impacts the performance of \project. In the rest of our benchmarks, we directly evaluate the performance of LSHS in comparison to related solutions by manually tuning block partitioning and node/worker/process grid layouts where applicable. We also evaluate \project's automatic partitioning heuristic on a data science problem in Section \ref{sec:data-science}.



\project\ is designed for medium-scale (terabytes) problem sizes. We evaluate on problem sizes of this scale in order to highlight the benefits \project\ does provide. We provide limitations to our approach in Section \ref{sec:analysis}, which show that \project's performance degrades as the number of array partitions increase, and when partition sizes are too small.

We run all CPU experiments on a cluster of 16 \texttt{r5.16xlarge} instances, each of which have 32 Intel \mbox{Skylake-SP} cores at 3.1Ghz with 512GB RAM connected over a 20Gbps network. Each AWS instance is running Ubuntu 18.04 configured with shared memory set to 512GB. For CPU-based experiments, the Ray cluster uses 312GB for the object store, and 200GB for workers. All \project\ experiments run on a single thread for BLAS operations and use only Ray worker nodes for computation, leaving the head node for system operations.

Unless otherwise noted, all experiments are executed by sampling data using random number generators, and all experiments are repeated 12 times. The best and worst performing trials are dropped to obtain better average performance. This is done primarily to avoid bias results due to cold starting benchmarks on Dask, Ray, and Spark.

We evaluate Dask's logistic regression (Dask ML 1.6.0) and QR decomposition, which implements the direct tall-skinny QR decomposition~\cite{tsqr}. 
We evaluate Spark-MLlib's (v2.4.7) logistic regression and QR decomposition. Logistic regression uses the L-BFGS solver from Breeze, which is an open source Scala library for numerical processing. Mllib's QR decomposition implements the indirect tall-skinny QR decomposition~\cite{tsqr-indirect} and uses the QR implementation from Breeze, which is internally implemented using LAPACK.

\subsection{Microbenchmarks}

\begin{figure*}
    \centering
    \includegraphics[width=0.95\textwidth]{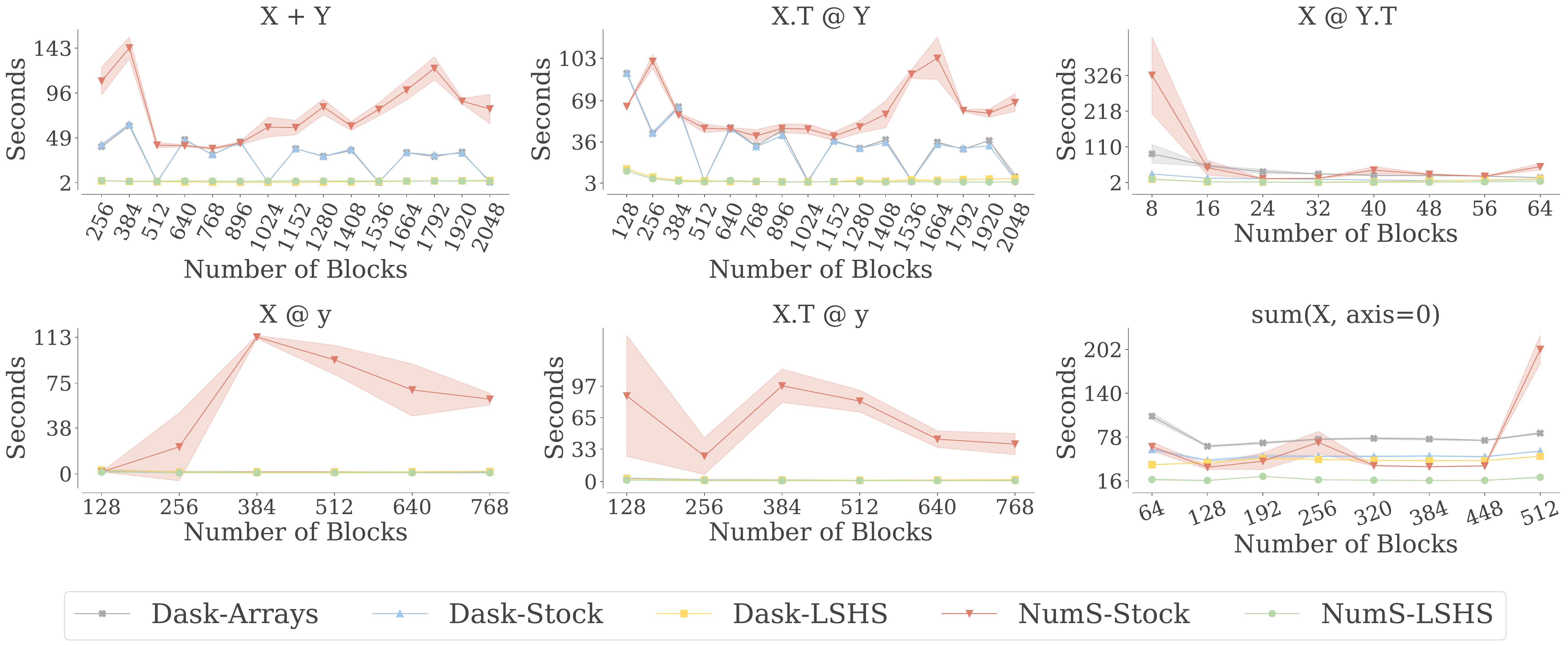}
    \vspace{-0em}
    \caption{
    An ablation study comparing \project\ on Dask and Ray, with and without LSHS, and include Dask Arrays as an additional point of comparison.
    All experiments are run on 16 node clusters, with 32 workers per node, for a total of 512 workers. In all but $\numssum$, $\X, \Y$ are 1 terabyte arrays, partitioned row-wise. $\y$ is partitioned to match the partitioning of $\X$ in $\X @ \y$ and $\X^\top @ \y$. $\numssum$ is executed on a multi-dimensional tensor partitioned along its first axis. \vspace{-0em}}
    \label{fig:micro-ablation}
\end{figure*}

The results of our ablation study are given in Figure \ref{fig:micro-ablation}. 
In every experiment, \project\ on Ray is significantly enhanced by LSHS. 
For $\X + \Y$ and $\X^\top @ \Y$, \project\ on Dask without LSHS and Dask Arrays achieve good performance whenever the number of partitions is divisible by the number of workers, whereas LSHS performs addition with $0$ and $\X @ \Y^\top$ with minimal communication.
With fewer partitions, we believe $\X @ \Y^\top$ is sub-optimal on Dask and Ray due to assigning large creation operations on fewer nodes, resulting in under-utilization of available cluster resources.
For matrix-vector operations, LSHS does not provide any significant enhancement over Dask scheduling. The optimal scheduling behavior is to move $\y$ to the nodes on which the partitions of $\X$ reside. For \project\ on Ray without LSHS, we observe object spilling due to too many large objects being assigned to a few nodes, and large object transmissions between nodes.
For $\numssum$, we measure the performance of reductions over large object transmissions.
Ray's high inter-node throughput achieves good results for this experiment. We believe Dask Array's poor performance on this task is due to a sub-optimal tree reduce, pairing partitions which are not co-located. 
Overall, we see that \project\ on Ray is the most robust to partitioning choices, and achieves well-rounded performance.
See Section \ref{sec:analysis} for our theoretical analysis, which supports the claims we make in these results.


\subsection{DGEMM}
\label{sec:dgemm}

\begin{table}[ht]
    \centering
    \begin{tabular}{|r|r|r|r|r|r|}
    \hline
        & {\bf 2GB} & {\bf 4GB} & {\bf 8GB} & {\bf 16GB} & {\bf 32GB}
        \\ \hline
        {\bf ScaLAPACK} & 224 & 480 & 992 & 64 & 992
        \\\hline
        {\bf SLATE} & 992 & 928 & 992 & 1408 & 992
        \\\hline
        {\bf \project} & 3953 & 3727 & 3953 & 5591 & 5271
        \\\hline
    \end{tabular}
    \caption{Square block size settings for ScaLAPACK, SLATE, and \project\ on the DGEMM benchmark. DGEMM is distributed over 1 node for 2GB, 2 nodes for 4GB, etc. up to 32GB on 16 nodes.}
    \label{fig:dgemm-table}
\end{table}
For dense square matrix multiplication, we compare to state-of-the-art baselines ScaLAPACK and SLATE~\cite{scalapack, slate}. We start with 2GB matrices on a single node, and double the amount of data as we double the number of nodes. We tune all libraries to their optimally performing block dimension. These settings are given in Table \ref{fig:dgemm-table}.

\begin{figure}[ht]
    \centering
        \includegraphics[width=0.6\columnwidth, trim=0.25cm 0.25cm 0.25cm 1.25cm, clip]{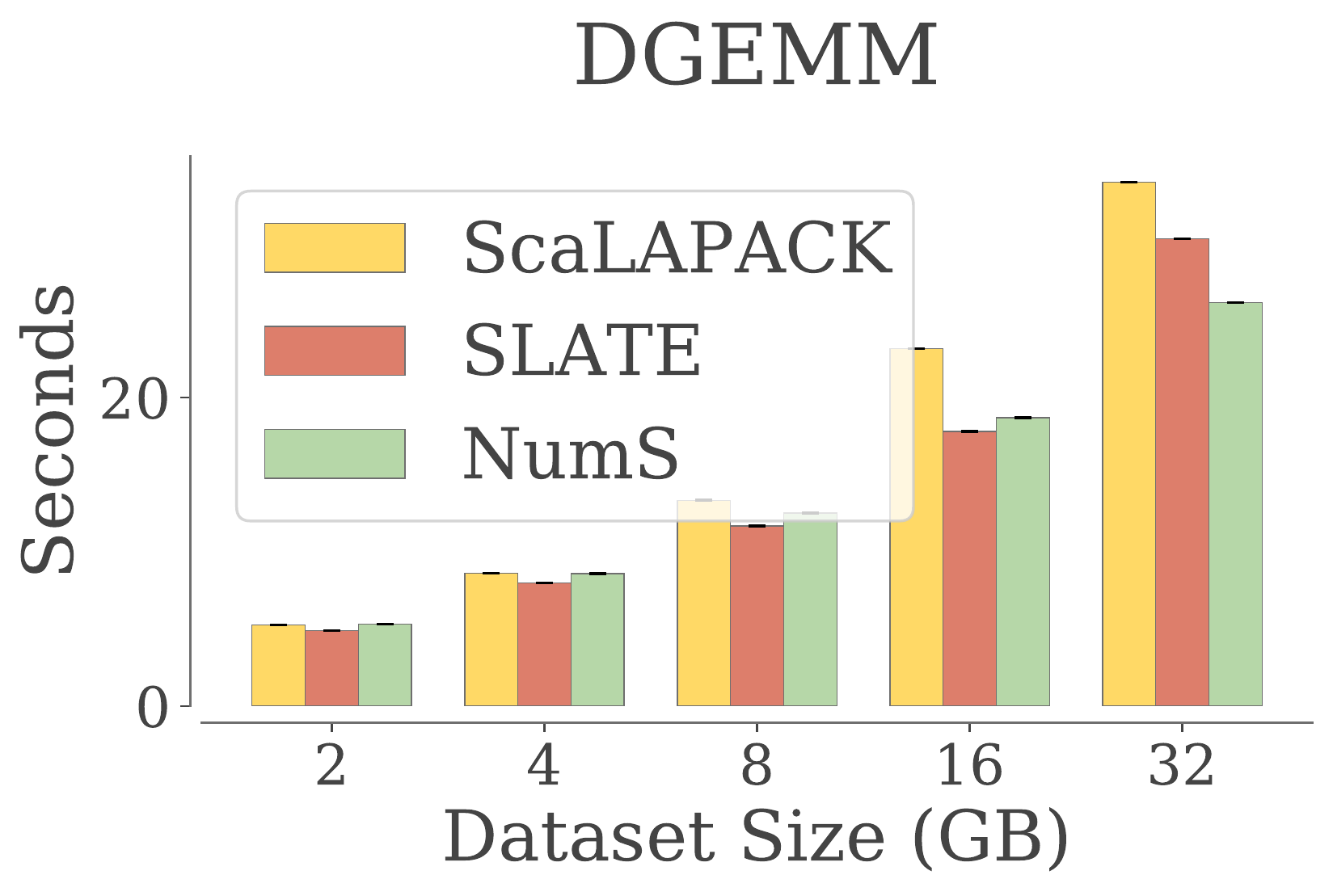}
        \vspace{-0em}
        \caption{Dense square matrix-matrix multiplication.}
        \vspace{-0em}
        \label{fig:matrix-multiply}
\end{figure}
Figure \ref{fig:matrix-multiply} shows that \project\ is competitive with HPC libraries on this benchmark. Both ScaLAPACK and SLATE implement the Scalable Matrix Multiplication Algorithm (SUMMA)~\cite{summa} for their dgemm routine.
While these results may seem surprising, the theoretical results we present in Section \ref{sec:analysis}, as well as our comprehensive analysis in Appendix \ref{appendix:analysis} show that our approach to parallelizing and scheduling distributed arrays can attain asymptotically lower communication time for this operation.
We show the existence of a square dense matrix-matrix multiplication algorithm which is asymptotically faster than SUMMA in $k$, suggesting that \project's performance on DGEMM improves as the number of nodes in a cluster increases.
While the SUMMA algorithm provides good communication bounds on distributed memory, it assumes every process has equivalent communication time. While the difference in communication time between vs. within nodes on supercomputers may not be significant, on multi-node clusters in the cloud, inter-node communication is generally much more expensive than intra-node communication.
LSHS places data and operations in a network topology-aware manner. For a particular operation, LSHS approximates the location of data, including data which is cached by Ray's object store from previous object transmissions. This enables LSHS a greater variety of operator placement decisions, enabling it to reduce inter-node communication by placing operations on nodes where operands are already co-located. Coupled with a locality-aware reduction operation, the final set of summations invoked in this operation can be performed entirely locally prior to performing an inter-node reduction. Furthermore, the overhead associated with performing intra-node communication among processes is implicit given Ray's shared-memory object store.

On the other hand, the presence of $\gamma$ limits \project's scalability in the limit of $k$. As we scale the size of data, we require more nodes and more partitions, which in turn puts greater demand on the number of RFCs which need to be dispatched by the driver process.
MPI-based libraries can be viewed as programs with distributed control, which provide a means to avoiding these issues.

\subsection{Linear Algebra}

QR decomposition is a core operation on a variety of linear algebra and data science operations, including linear regression, singular value decomposition, and principal component analysis~\cite{tsqr, tsqr-indirect, bishop}. In this section, we evaluate the weak scaling of \project\ on direct and indirect QR decomposition, as well as a performance comparison to Dask, and Spark. We also include results for \project\ without LSHS to highlight the role of scheduling. All experiments perform the same number of steps and operations. Each experiment is repeated 12 times, and the best and worst performing trials are dropped to avoid bias due to cold starts.

The QR decomposition of a matrix $A$ finds the matrices $Q$ and $R$ such that $A = QR$. A direct QR decomposition computes $Q$ from intermediate factors of $Q$, whereas indirect QR discards intermediate $Q$ factors and computes $Q$ via the operation $A R^{-1}$.

\begin{figure}[t]
    \centering
    \begin{subfigure}[b]{0.4\columnwidth}
      \centering
      \includegraphics[width=\textwidth, trim=0.25cm 0.25cm 0.25cm 1.25cm, clip]{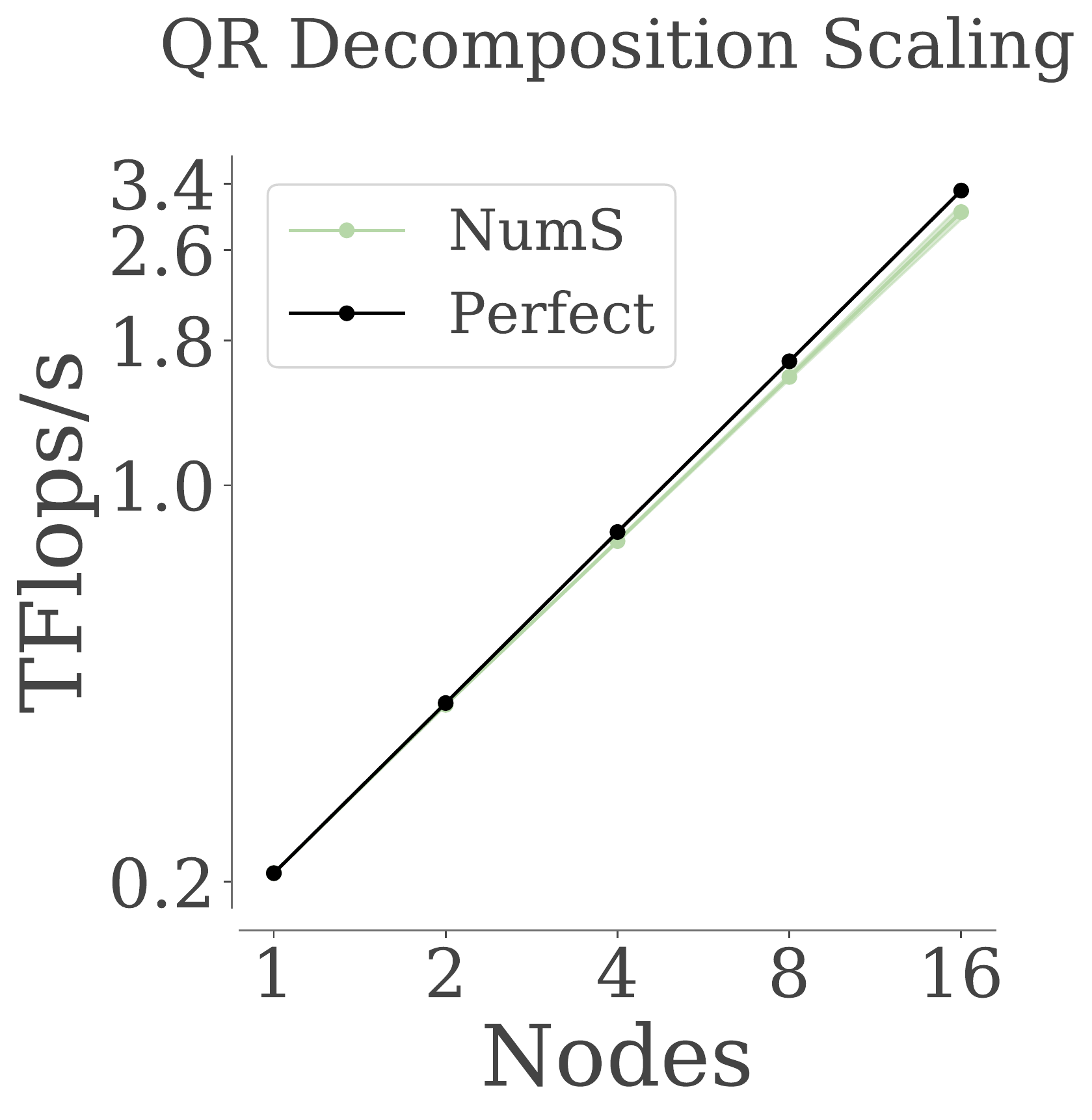}
      \caption{QR Decomposition.}
      \vspace{-0em}
      \label{fig:qr-scaling}
    \end{subfigure}
    ~
    \begin{subfigure}[b]{0.4\columnwidth}
      \centering
      \includegraphics[width=\textwidth, trim=0.25cm 0.25cm 0.25cm 1.25cm, clip]{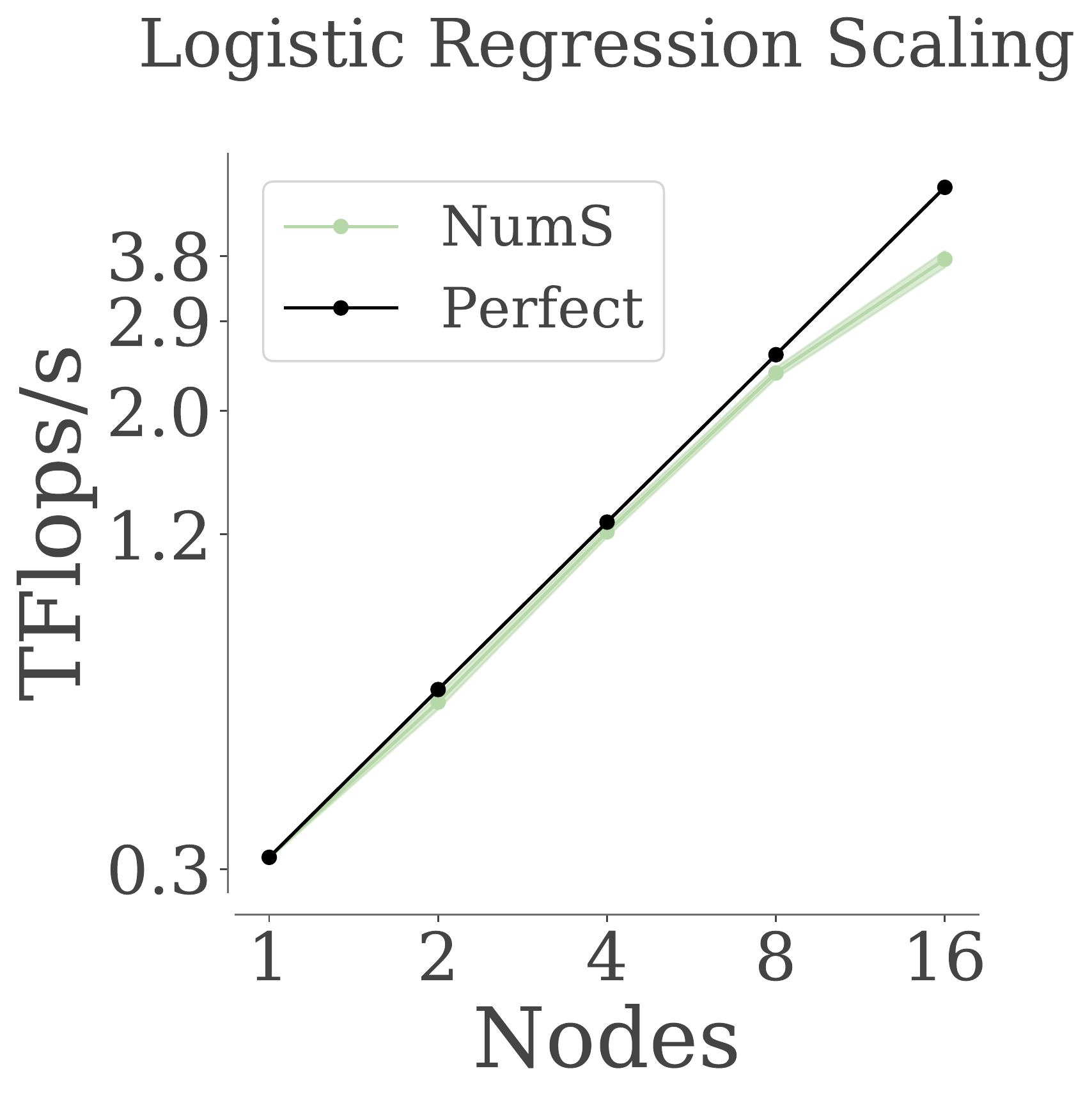}
      \caption{Logistic Regression.}
      \vspace{-0em}
      \label{fig:logistic-scaling}
    \end{subfigure}
    \caption{QR decomposition achieves near-perfect scaling. Logistic regression exhibits a slowdown at 16 nodes due intermediate reduction operations over a 20Gbps network.}
    \vspace{-0em}
    \label{fig:scaling}
\end{figure}

The weak scaling of indirect QR decomposition is carried out by doubling the amount of resources as we double the amount of work, starting with 64GB of data on a single node. Scaling is near perfect (Figure \ref{fig:qr-scaling}). 

\begin{figure}[ht]
    \centering
    \begin{subfigure}[b]{0.45\columnwidth}
      \centering
      \includegraphics[width=\textwidth, , trim=0cm 0cm 0cm 1.2cm, clip]{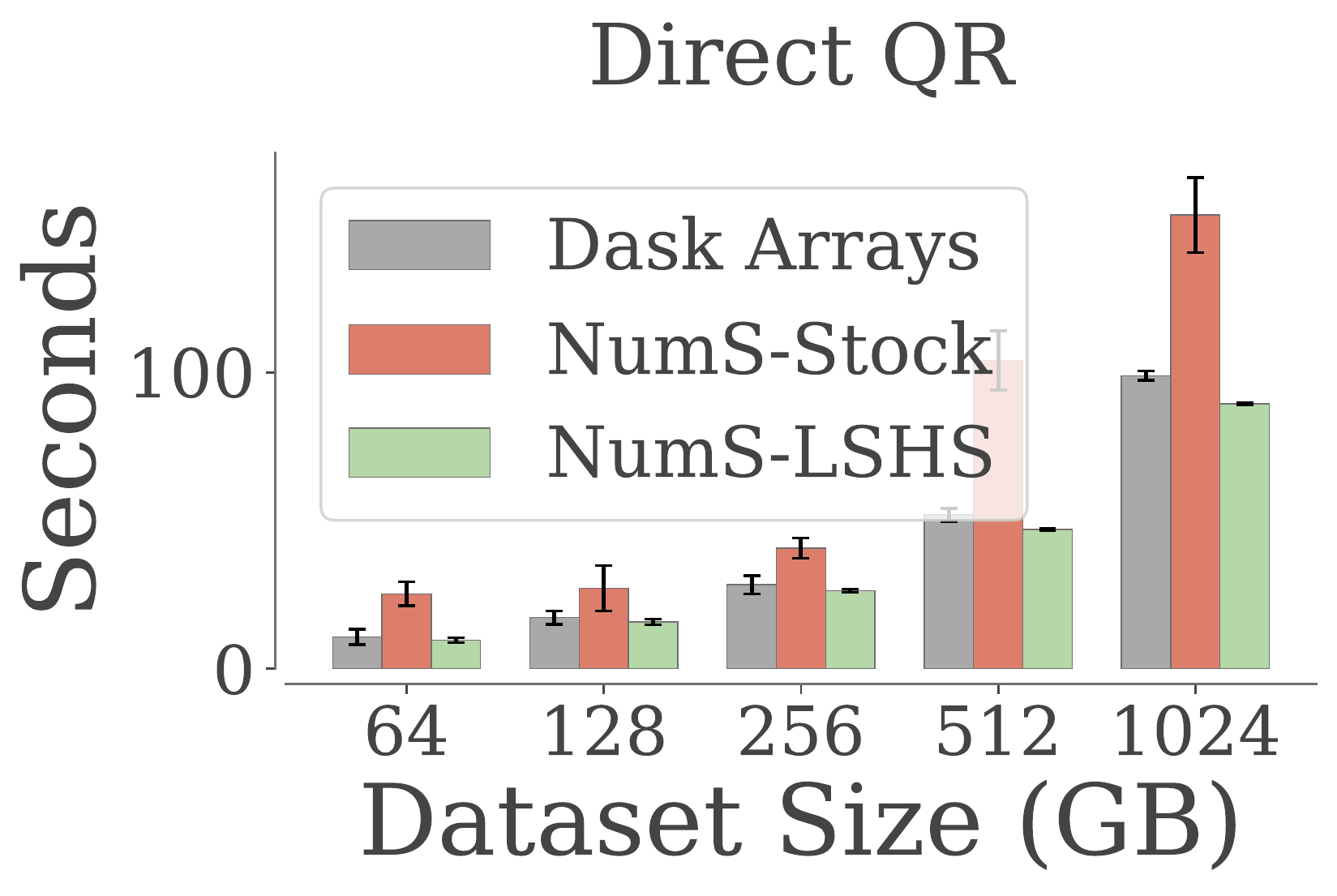}
      \caption{Direct TSQR.}
      \vspace{-0em}
      \label{fig:qr-direct}
    \end{subfigure}
    ~
    \begin{subfigure}[b]{0.45\columnwidth}
      \centering
      \includegraphics[width=\textwidth, , trim=0cm 0cm 0cm 1.2cm, clip]{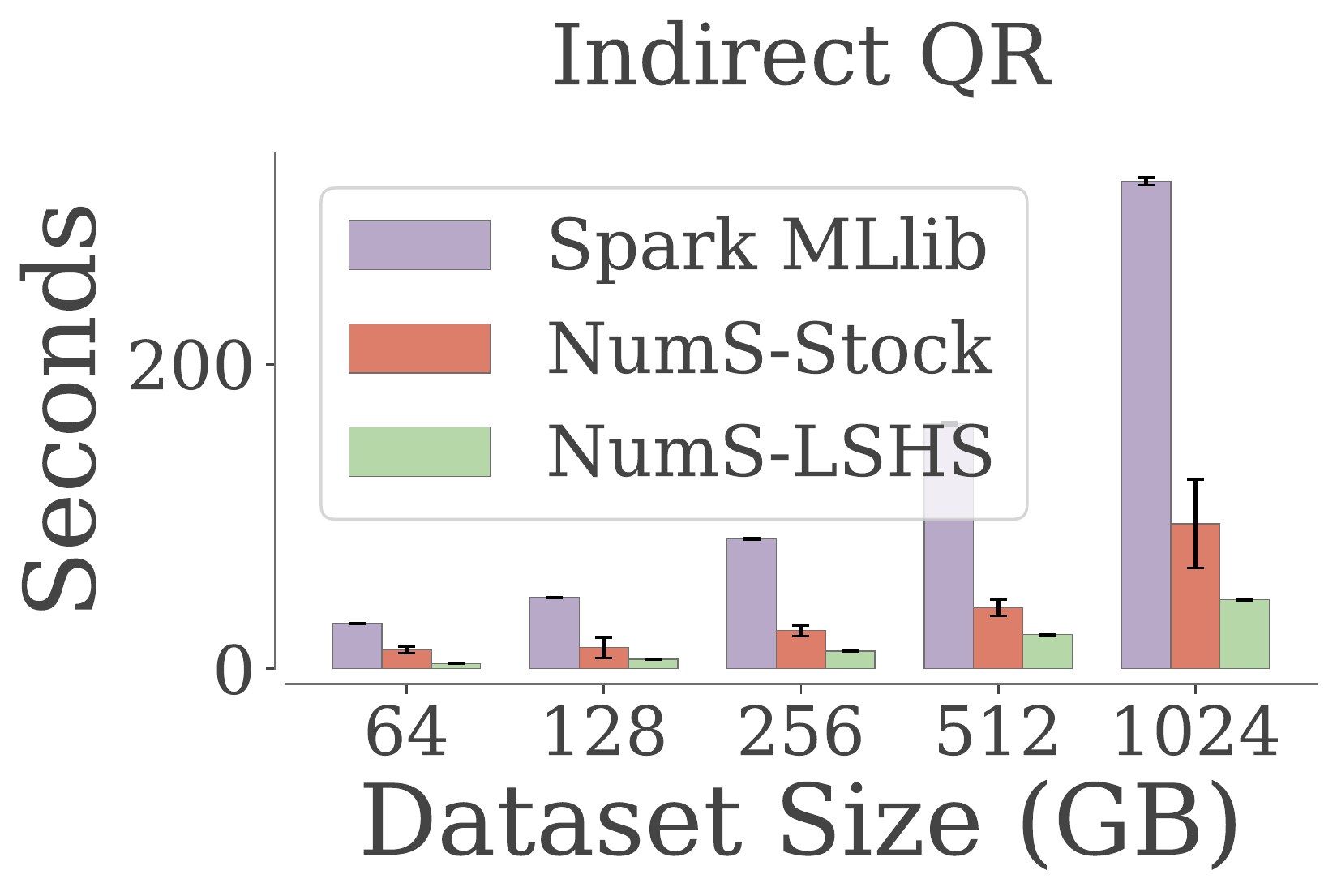}
      \caption{Indirect TSQR.}
      \vspace{-0em}
      \label{fig:qr-indirect}
    \end{subfigure}
    \caption{TSQR execution time on \project, Dask, and Spark.}
    \vspace{-0em}
    \label{fig:qr}
\end{figure}

We compare Dask and \project\ on the direct tall-skinny QR decomposition algorithm~\cite{tsqr}. We sample data row-wise in $2$GB blocks, which achieves peak performance for both libraries.
Figure \ref{fig:qr-direct} shows the results of our direct TSQR benchmarks.
\project\ performs comparably to Dask on this benchmark. Dask's direct tall-skinny implementation requires a single column partition, leaving only one dimension along which data is partitioned. The partitioning of data for Dask which achieves peak performance implicitly results in data locality for a number of intermediate operations due to Dask's round-robin placement of initial tasks. We observe this behavior directly in our ablation study in Figure \ref{fig:micro-ablation}.

Since Spark does not implement a direct QR decomposition, we evaluate both \project\ and Spark on indirect TSQR decomposition. Similar to logistic regression, Spark's implementation of indirect TSQR decomposition is sensitive to partition tuning. 
Figure \ref{fig:qr-indirect} compares the results of our indirect TSQR implementation to Spark's. Similar to direct TSQR for Dask, indirect TSQR is a statically scheduled algorithm for Spark and \project. Since both algorithms are identical, and scheduling is static, we attribute the difference in performance to differences between Spark and Ray.

\subsection{Tensor Algebra}
\label{sec:eval-tensor}

\begin{figure}[t]
    \centering
    \begin{subfigure}[b]{0.45\columnwidth}
      \centering
      \includegraphics[width=\textwidth, trim=0cm 0cm 0cm 1.2cm, clip]{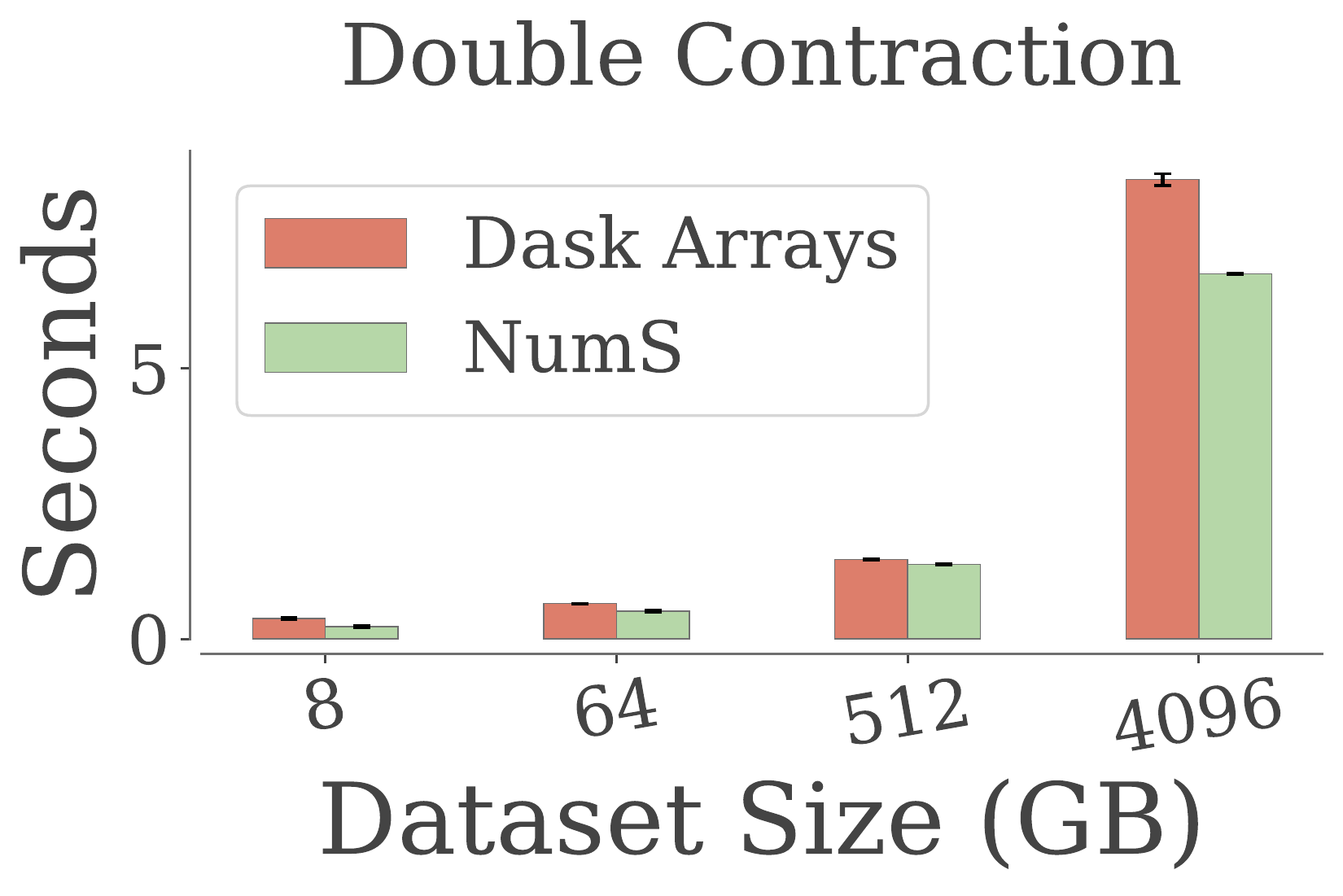}
      \caption{Double Contraction.}
      \vspace{-0em}
      \label{fig:double-contraction}
    \end{subfigure}
    ~
    \begin{subfigure}[b]{0.45\columnwidth}
      \centering
      \includegraphics[width=\textwidth, trim=0cm 0cm 0cm 1.2cm, clip]{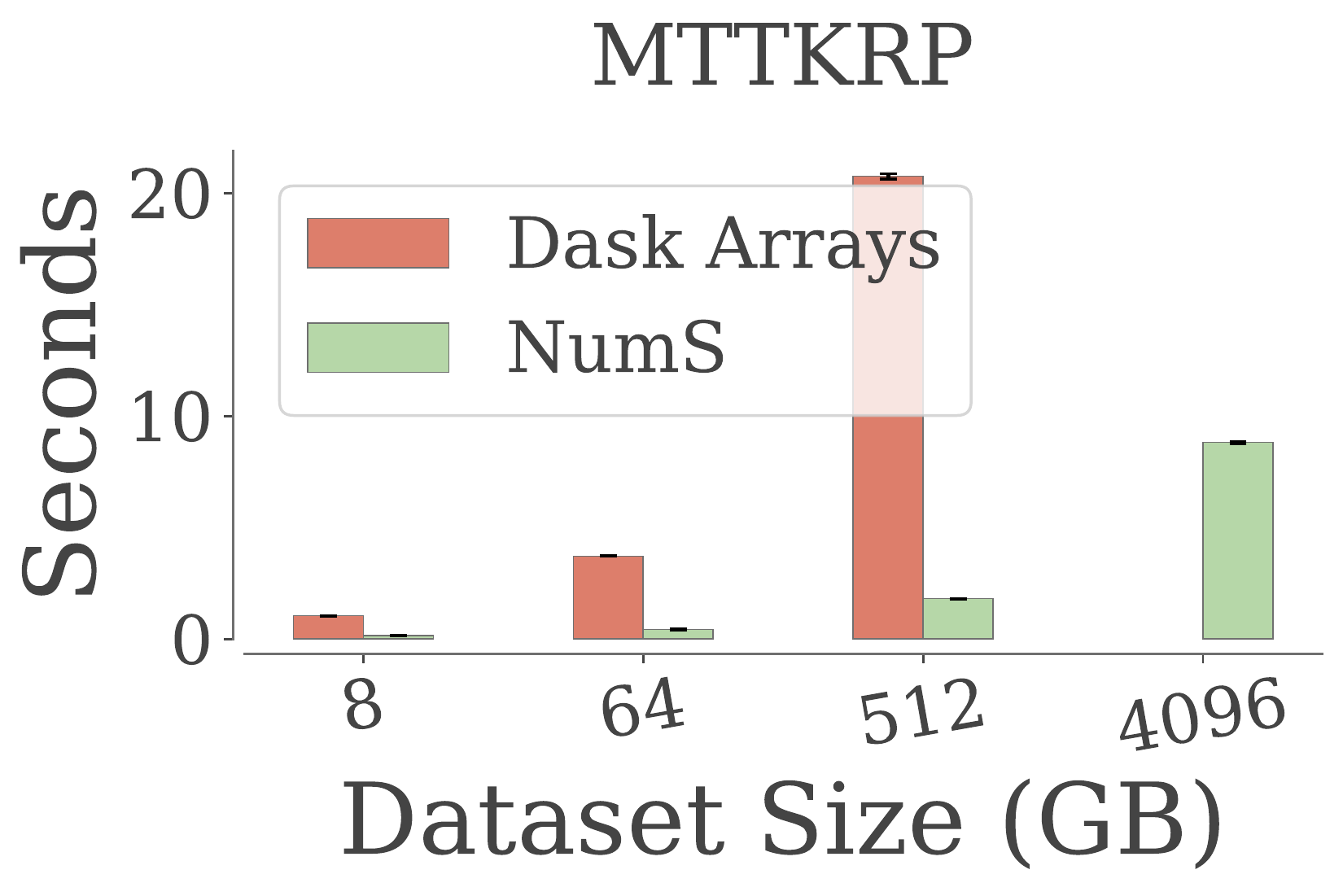}
      \caption{MTTKRP.}
      \vspace{-0em}
      \label{fig:mttkrp}
    \end{subfigure}
    \caption{Tensor algebra on \project\ and Dask Arrays.}
    \vspace{-0em}
    \label{fig:tensor}
\end{figure}

We compare \project\ to Dask Array's implementation of the $\tensordot$ and $\einsum$ operators, primitives which enable the expression of distributed dense tensor algebra operations. We perform these operations on a 16 node cluster with 32 workers per node.
For the $\einsum$ operator, we perform the Matricized Tensor Times Khatri Rao Product (MTTKRP), which we express in Einstein summation notation. This operation is the closed-form solution to the alternating least squares algorithm for  tensor factorization \cite{als-tf}. For the $\tensordot$ operator, we perform the standard tensor double contraction on operands which frequently occur in a variety of other tensor decompositions\cite{tensor-decomp}.

For MTTKRP, we sample $\X \in \mathbb{R}^{I \times J \times K}$, $\B \in \mathbb{R}^{I \times F}$, $\B \in \mathbb{R}^{J \times F}$, and perform $\einsum(ijk,if,jf->if, \X, \B, \C)$. For both Dask and \project, we partition every array to achieve peak performance. In \project, we also tune a cubic configuration of the available compute nodes, to further control the mapping of partitions to nodes. We set $F=100$ and vary the number of elements of $I=J=K$ to set the size of $\X$ from $8$GB to $4$TB.

For tensor double contraction, we sample $\X$ identically to the MTTKRP benchmark, and sample $\Y  \in \mathbb{R}^{J \times K \times F}$ with $F=100$. We also vary the size of $\X$ identically to the MTTKRP benchmark.

Both results depend heavily on LSHS. Array partitioning and node grid also play a significant role. For \project, MTTKRP partitioned along dimension $J$, and a node grid of $16 \times 1 \times 1$ performed best.
For the double contraction benchmark, a node grid of $1 \times 16 \times 1$ performed best, with relatively balanced partitioning along dimensions $J$ and $K$.
In both benchmarks, both \project\ and Dask Arrays perform a collection of sum-of-products. Both libraries perform tree-based reductions. 

For MTTKRP, Dask performed best with relatively balanced partitioning of blocks, but is unable to achieve good initial data placement to minimize inter-node communication due to its inability to specify a node grid. Furthermore, its reduction tree is constructed before any information about the physical mapping of blocks to nodes is available, resulting summations between vs. within nodes. After tuning, the $4$TB benchmark on Dask Arrays took approximately $241.9$ seconds (20$\times$ slower than \project, which is excluded from Figure \ref{fig:tensor}.

For double contraction, Dask and \project\ performance is relatively the same. Unlike MTTKRP, there is no good factoring of 16 nodes into a node grid that LSHS can take advantage of in order to reduce inter-node communication. This is mainly due to the structure of the problem: The tensor contraction sums over the dimensions $J$ and $K$, and the ordering of dimensions for tensors $X$ and $Y$ align only along dimension $J$. This is why the $1 \times 16 \times 1$ factoring of the nodes performs best.


\subsection{Generalized Linear Models}
\label{sec:glm-eval}

Generalized linear models (GLMs) are notoriously difficult to scale due to their reliance on basic array and linear algebra operations. Furthermore, optimizing these models with second order methods require the expensive computation of the Hessian matrix, or approximations to the Hessian matrix \cite{wright, bishop}. Theoretically, Newton's method converges faster than any other method available for GLMs~\cite{wright}. In practice, fast convergence can be difficult to achieve without proper utilization of all computational resources when fitting these models to large amounts of data. \project\ is able to achieve high performance on any model which relies heavily on element-wise and basic linear algebra operations, making GLMs an ideal \project\ application. 
We explain in greater detail the execution of GLMs in Section \ref{sec:glm}.

Since logistic regression is the most widely used GLM, we evaluate our implementation of logistic regression. We measure its weak scaling performance in terms of teraflops, measure its network and memory load with and without LSHS, and compare its performance to other solutions.
Our experiments are executed on synthetic classification data. Our data is drawn from a bimodal Gaussian with $75\%$ of the data concentrated at mean 10 with variance 2 (negative samples), and the remaining $25\%$ concentrated at mean 30 with variance 4 (positive samples). Each sample is 256-dimensional. We sample from these distributions to satisfy the required dataset size. For example, a 64GB dataset of 64bit floats consists of a design matrix $X$ with $31,250,000$ rows and $256$ columns, and a target vector $y$ consisting of $31,250,000$ values $\in \{0,1\}$.
This data size and distribution was recommended by our industry collaborators.

Our weak scaling results for logistic regression are near perfect until 16 nodes, at which point performance degrades due to inter-node reductions over a 20Gbps network (Figure \ref{fig:logistic-scaling}).

We compare \project's performance on logistic regression to Dask ML and Spark MLlib. We include results for \project\ on Ray without LSHS to highlight the role of scheduling.
Dask and Spark implement different versions of these algorithms, so we implement both versions of both algorithms for a fair comparison.
In these experiments, we hold the cluster resources fixed at 16 nodes, varying only the dataset size to evaluate the performance of each system.
All experiments perform the same number of steps and operations.

\begin{figure}[t]
    \centering
    \begin{subfigure}[b]{0.45\columnwidth}
      \centering
      \includegraphics[width=\textwidth, trim=0cm 0cm 0cm 1.2cm, clip]{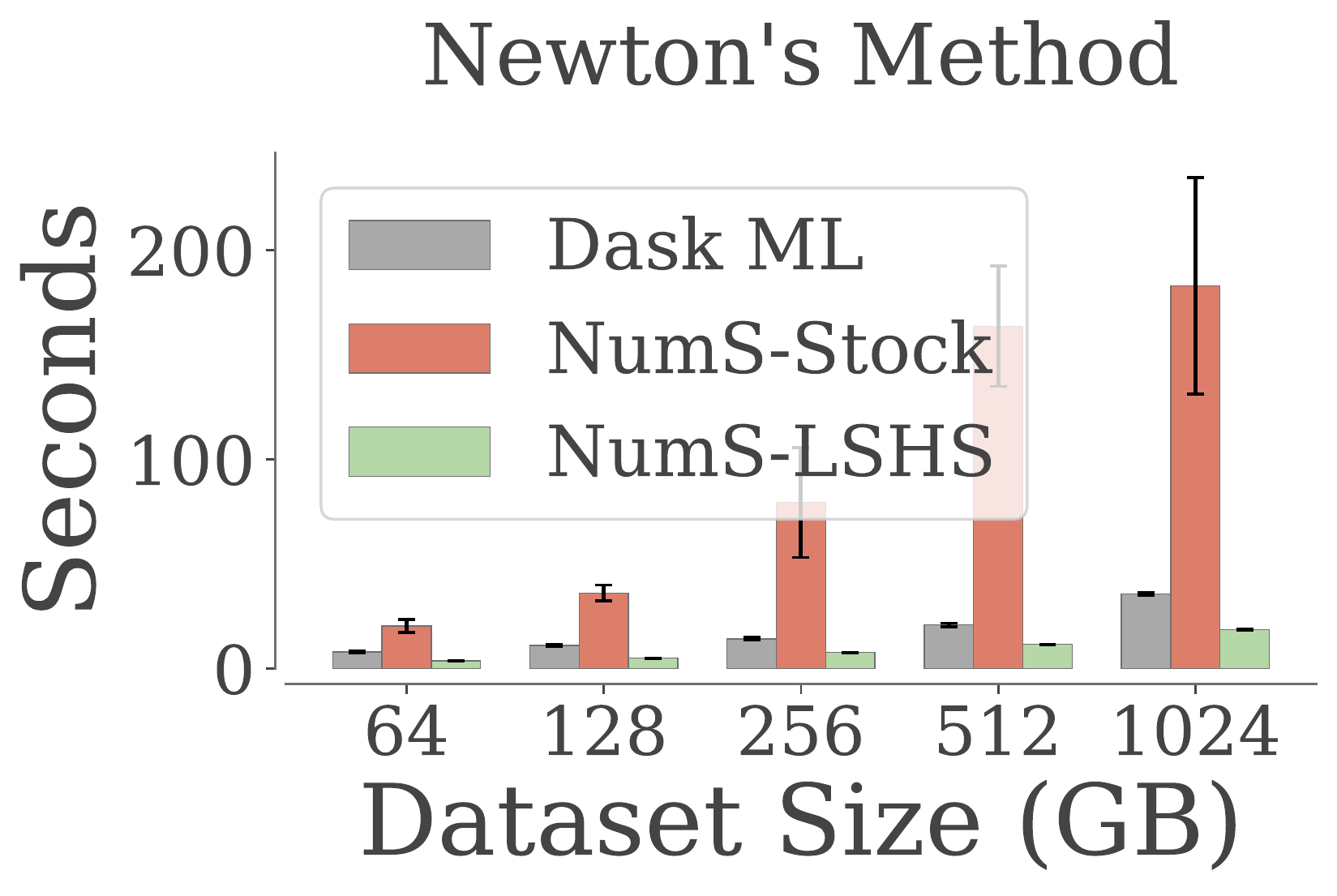}
      \caption{Newton's method.}
      \label{fig:logistic-newton}
    \end{subfigure}
    ~
    \begin{subfigure}[b]{0.45\columnwidth}
      \centering
      \includegraphics[width=\textwidth, trim=0cm 0cm 0cm 1.2cm, clip]{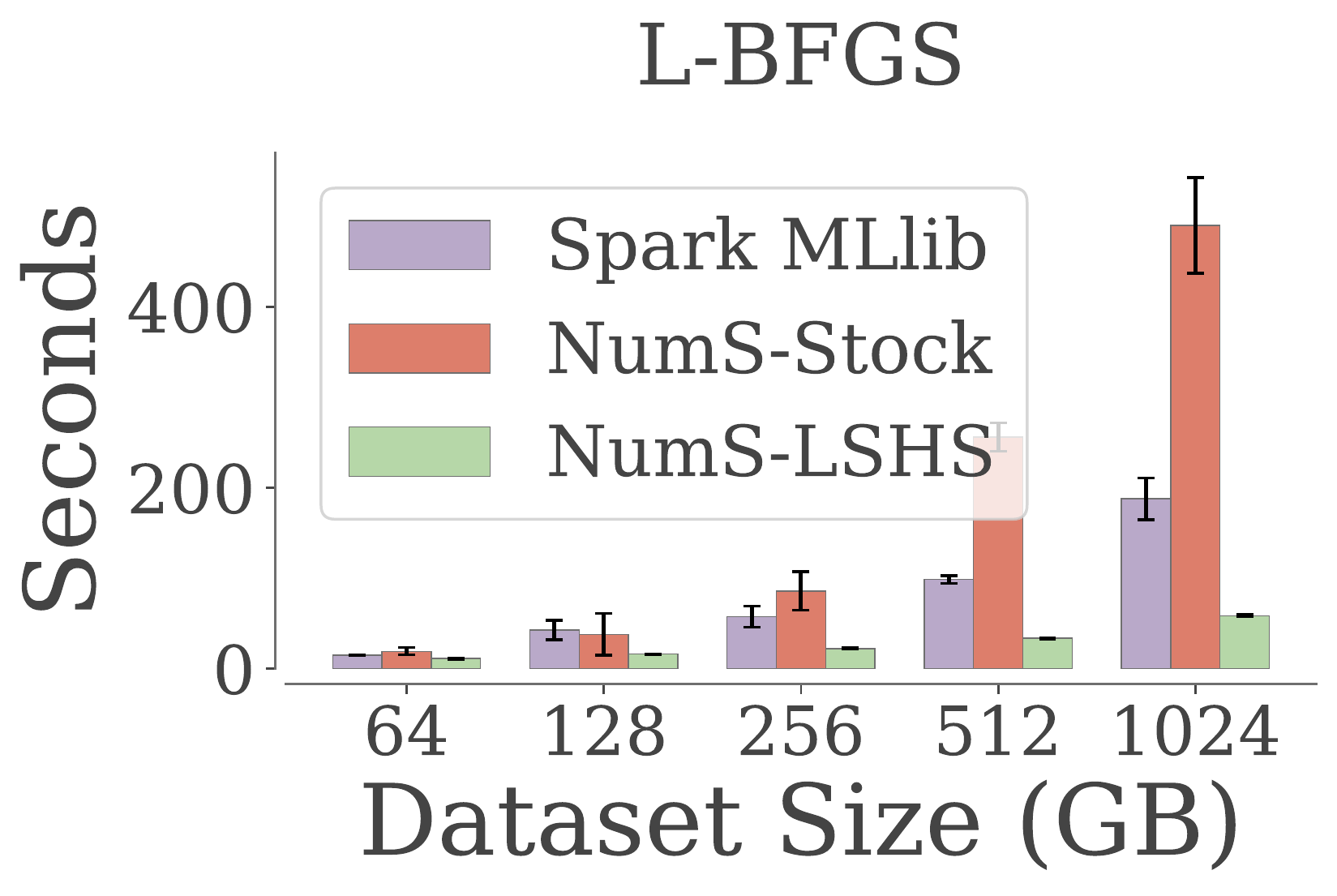}
      \caption{L-BFGS.}
      \label{fig:logistic-lbfgs}
    \end{subfigure}
    \vspace{-0em}
    \caption{Logistic regression fitting time.}
    \vspace{-0em}
    \label{fig:logistic}
\end{figure}

For our comparison to Dask, we sample data row-wise in $2$GB blocks, which yields peak performance for both Dask and \project. We use Newton's method for both libraries.
Newton's method is optimal for logistic regression's convex objective.
Figure~\ref{fig:logistic-newton} shows that \project\ outperforms Dask at every dataset size. The performance gap is partially explained by differences between LSHS and Dask's dynamic scheduler. We believe the majority of the performance gap is due to Dask ML's implementation logistic regression which, based on our inspection of their source code, aggregates gradient and hessian computations on the driver process to perform updates to model parameters and test for convergence.


Since Spark does not support Newton's method, we compare our implementation of the L-BFGS optimizer to Spark's version. We initialize our logistic regression implementation to execute 10 optimization steps, with no regularizer, and L-BFGS configured to use a history length of 10. Both implementations use identical line search algorithms and are configured identically. Figure~\ref{fig:logistic-lbfgs} shows that our implementation of logistic regression with the L-BFGS optimizer outperforms Spark. Spark's logistic regression and Breeze's L-BFGS ~\cite{breeze} implementation is a statically scheduled implementation. To our knowledge, the algorithms and scheduling of operations on partitions is identical to \project's implementation, and the scheduling behavior of LSHS for this problem. While LSHS is essential to achieve efficiency from a scheduling point-of-view, we believe the performance gap beyond scheduling is explained by differences between Spark and Ray.

To better understand why LSHS enhances \project's performance on Ray,
We measure memory, network in, and network out of every node at equally spaced intervals for \project\ on Ray with and without LSHS. For these experiments, we use 16 nodes with 32 workers per node and measure execution time of a single iteration of Newton's method on a 128GB logistic regression problem. 

These experiments measure execution time and resource utilization required to load data from S3, execute a single iteration of Newton's method. 
\begin{figure}[ht]
    \centering
    \includegraphics[width=.9\columnwidth]{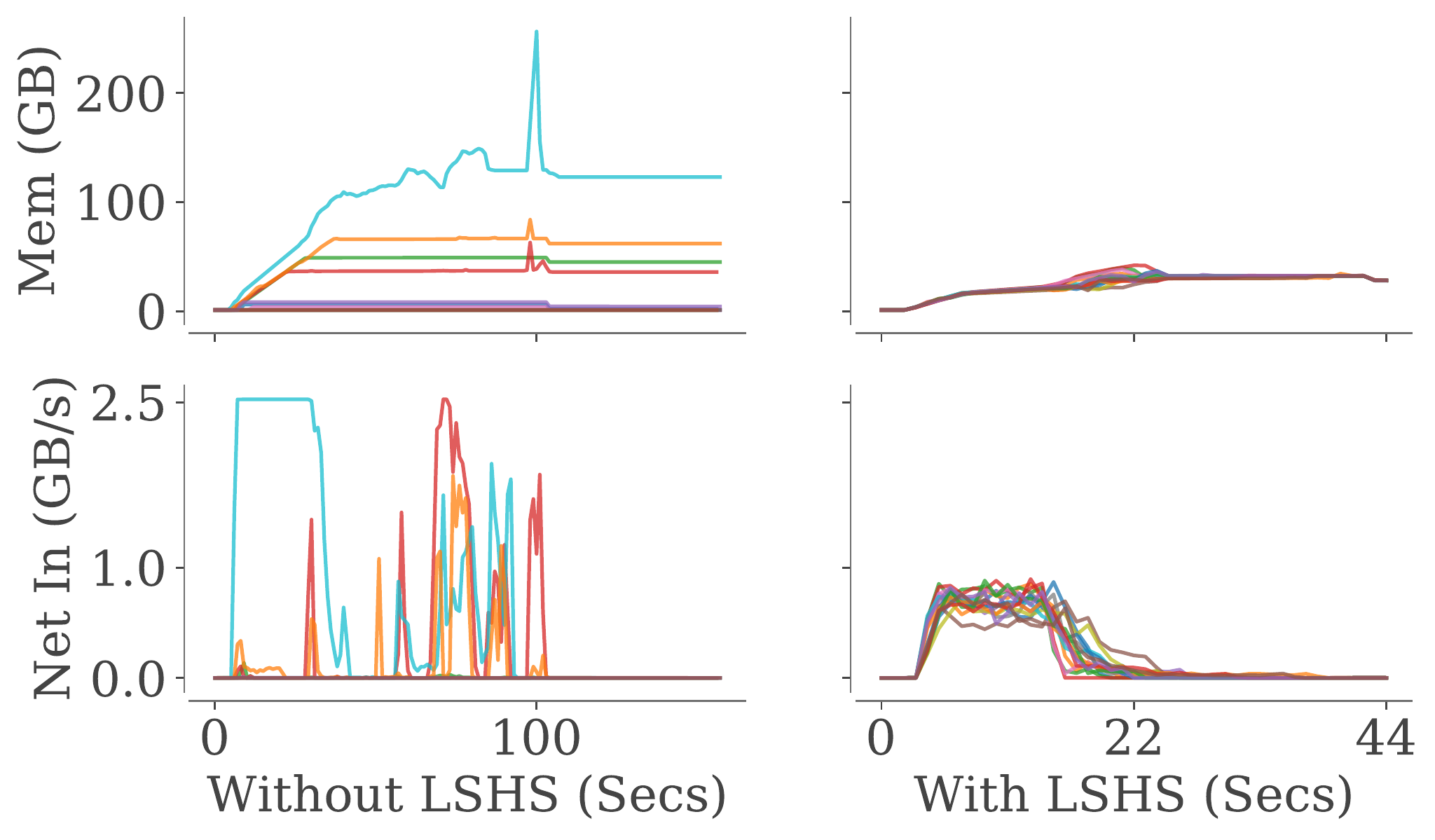}
    \vspace{-0em}
    \caption{Ablation on memory and network load.}
    \vspace{-0em}
    \label{fig:ablation-load}
\end{figure}

Figure \ref{fig:ablation-load} shows memory usage and network input over one iteration of Newton's method. Each curve tracks the load on one node. Densely clustered curves, where all nodes have similar load over the experiment, indicate good load balance. Lower y-axis values indicate lower resource footprint. LSHS significantly enhances \project\ performance on Ray in terms of load balance and resource footprint. Without LSHS, Ray executes the majority of submitted tasks on a single node, while LSHS distributes load without increased network communication. In particular, Ray's bottom-up scheduling does not define an explicit strategy for scheduling independent tasks \cite{ray}, resulting in a sub-optimal data layout for both element-wise and linear algebra operations. For this task, LSHS decreases network load by a factor of 2$\times$, uses 4$\times$ less memory, and decreases execution time by a factor of 10$\times$.

\subsection{Data Science}
\label{sec:data-science}

\project\ not only provides a speedup in distributed memory settings, but it also provides a significant speedup on a single node for certain data science applications. 
The experiments in this section make use of \project's automatic block partitioning (see Section \ref{sec:grapharray}), demonstrating that a speedup can be achieved by simply replacing Python import statements of the libraries used in this section with \project\ equivalents. While these experiments run on the same instances we have been using throughout this section, \project\ is pip-installable on your laptop and can provide comparable speedups to what we present in this section.



\begin{table}[ht]
    \centering
    \begin{tabular}{r|r|r|r|r}
        \hline
        \multicolumn{1}{c}{\textbf{Tool Stack}} & \multicolumn{1}{c}{\textbf{Load}} & \multicolumn{1}{c}{\textbf{Train}} &
        \multicolumn{1}{c}{\textbf{Predict}} & \multicolumn{1}{c}{\textbf{Total}}
        \\ \hline
        Python Stack & 65.55 & 61 & 0.43 & 126.98
        \\\hline
        \project\ & 11.79 & 3.21 & 0.20 & 15.2
        \\\hline
    \end{tabular}
    \caption{\project\ vs. a Python stack consisting of Pandas for a data loading, and scikit-learn on NumPy for training a logistic regression model. All values are reported in seconds.}
    \vspace{-0.0em}
    \label{fig:speedup-sklearn}
\end{table}

Pandas, NumPy, and scikit-learn make up a common stack for data science in Python. \project\ provides a parallel \texttt{read\_csv} method comparable to Pandas', eliminating one layer in the package stack for numerical CSV files. Table \ref{fig:speedup-sklearn} shows that \project\ achieves an 8$\times$ speedup over Pandas'~\cite{pandas} serial \verb|read_csv| operation with scikit-learn's training and prediction procedures for a logistic regression model trained on the 7.5GB HIGGS dataset~\cite{higgs}. Both \project\ and scikit-learn's logistic regression are configured to use 32 cores. 

We tune scikit-learn to use its fastest optimizer, which is l-bfgs. Compared to \project' Newton optimizer, l-bfgs requires significantly more iterations to converge. Furthermore, l-bfgs requires line search at every iteration, which requires multiple calls to the logistic regression objective function.
Newton's method does not require a line search and converges in fewer iterations than l-bfgs. Our Newton's method implementation is optimized for the kinds of tall-skinny matrices which commonly occur in data science, achieving greater utilization of available memory and cores through efficient parallelization of basic linear algebra operations.

\begin{figure}[ht]
    \centering
    \includegraphics[width=0.6\columnwidth]{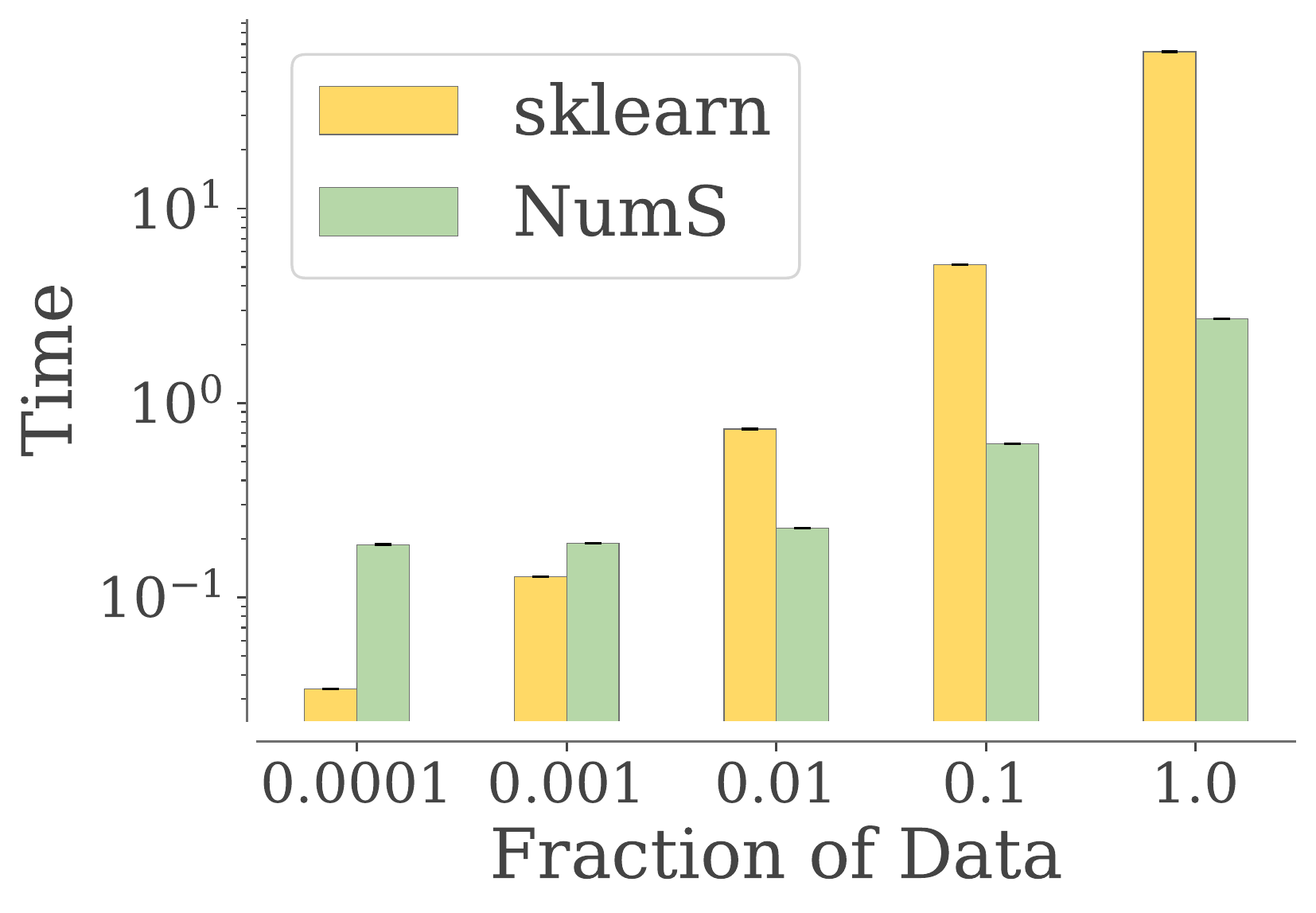}
    \vspace{-0em}
    \caption{Training on fractions of the HIGGS dataset.}
    \vspace{-0em}
    \label{fig:sklearn-nums-scale}
\end{figure}

To further dig into these differences, we compare \project\ to scikit-learn on different fractions of the HIGGS dataset. Figure \ref{fig:sklearn-nums-scale} shows that, at smaller scales, \project\ is $5\times$ slower than scikit-learn, and at larger scales, it is $20\times$ faster than scikit-learn.

Beyond differences in the optimizer, the primary difference in performance is due to \project's parallelization of all array operations, not just those parallelized by the underlying system's BLAS implementation. We measure this by implementing Newton's method in pure NumPy, with full parallelization of BLAS operations (32 cores). We measure the amount of time spent in serial operations vs. parallel operations, and we find that \textit{$90$\% of the time for Newton's method using NumPy is spent on serial operations}. In total, our NumPy implementation of Newton's method takes approximately $11$ seconds, a $5\times$ speedup over scikit-learn's fastest optimizer. Compared to this implementation of Newton's method, \project\ achieves a speedup of $3.5\times$.
Compared to Newton's method, l-bfgs is comprised of less expensive matrix operations (no direct computation of the Hessian), and many more serially executing element-wise operations.






\section{Conclusion And Future Work}
\label{sec:conclusion}
Our results show that \project\ achieves competitive performance with state-of-the-art HPC libraries, and in many instances outperforms related solutions.
Our theoretical and empirical analyses suggest that \project\ performs best on data on the order of gigabytes and terabytes. For smaller data, \project\ may be slower than traditional Python tools. While \project\ has not been evaluated on very large data (petabytes to exabytes), our theoretical analysis suggests systems such as SLATE are better suited for linear algebra operations at these scales.
Based on our work, we believe that all distributed data structures, not just distributed arrays, that rely on dynamic scheduling require some combination of LSHS with data layouts optimized for operations on those data structures. In particular, Dask's dataframes and project Modin~\cite{modin}, a portable dataframe abstraction that runs on Dask and Ray, could benefit from similar techniques presented in this work.
Future directions for \project\ include (1) generalizing our findings and providing a framework on which any distributed data structure can benefit from LSHS; (2) improving the usability of \project's user API by enhancing automatic block-partitioning and eliminating the need for a user-specified node grid; and (3) reducing RFC overhead by introducing operator fusion.

\section*{Acknowledgments}
\label{sec:ack}
Thank you to Amazon Core AI and Microsoft for supporting the evaluation of \project\ by providing AWS and Azure credits. 

\bibliographystyle{plain}
\bibliography{main}

\appendix


\section{Communication Analysis of LSHS}
\label{appendix:analysis}

We extend the $\alpha-\beta$ model of communication to analyze the communication time of element-wise, reduction, and basic linear and tensor algebra operations. In our model, for a particular channel of communication, $\alpha$ denotes latency and $\beta$ denotes the inverse bandwidth of the channel. We also model the time to dispatch an operation from the driver process as $\gamma$, called the \textit{dispatch latency}. The time to transmit $n$ bytes is given by $\alpha + \beta n$.

For a dense array of size $N$, let $p$ the number of workers, $N/p = n$ the block size, or number of elements, and $r = p/k$ the number of workers-per-node on a $k$ node cluster.

Let $C(n) = \alpha + \beta n$ denote the time to transmit $n$ bytes between two nodes within a multi-node cluster.
Let $D(n) = \alpha'' + \beta'' n$. This is the cost of transmitting data between workers within a single node in Dask. Let $R(n) = \alpha' + \beta' n$. This expression captures the implicit cost of communication between workers within a single node of Ray. Ray maintains a shared memory key/value store on each node, enabling every worker to access data written by any other worker without explicit communication.

We assume $\alpha >> \alpha'' > \alpha'$, and $\beta >> \beta'' > \beta'$. We expect $\alpha'' > \alpha'$ and $\alpha'' > \alpha'$ because Dask relies on the TCP protocol for object transmission between workers within the same node, whereas Ray writes data directly to Linux shared memory.

All lower bounds are given in terms of Ray's communication time. Our lower bounds depend on the assumption that a block need only be transmitted to a node once, after which it is cached by Ray's object store. We also assume Ray's object store is large enough to hold all intermediate objects which it caches.
We also assume that bytes may be sent and received in parallel, which is possible in Ray.
We present the $\gamma$ term once as part of the lower bound, since in all cases, the same number of operations are dispatched from the driver process.

\begin{algorithm}[ht]
\SetKwFunction{Reduce}{Reduce}
\SetKwFunction{RMM}{RMM}
\SetKwFunction{MATMUL}{MATMUL}
\SetKwFunction{NOTPARTITIONED}{NOTPARTITIONED}
\SetKwProg{Pn}{Function}{:}{\KwRet}
\Pn{\RMM{$\A$, $\B$}}{
    \uIf{\NOTPARTITIONED{$\A$, $\B$}}{
        \KwRet \MATMUL{$\A$, $\B$} \;
    }
    $\forall_{i=0, j=0, h=0}^{m-1, n-1, k-1} \, \{ \C'_{i,j,h} \gets \RMM{\A_{i,h}, \B_{h,j}} \}$ \;
    $\forall_{i=0, j=0}^{m-1, n-1}$ \{ $\C_{i, j} \gets$ \Reduce{$\C'_{i,j}$} \} \;
    \KwRet $\C$ \;
}
\caption{Recursive Matrix Multiplication.}
\label{algo:rmm}
\end{algorithm}
\vspace{-0em}
Our results are more easily presented and understood in terms of the recursive matrix multiplication algorithm, which is given by Algorithm \ref{algo:rmm}. Nested blocks of a recursively partitioned matrix $\X$ are obtained by the subscripts $((\X_{i_1, j_1})_{i_2, j_2})\dots)_{i_d, j_d}$, where $d$ is the depth of the nested blocks of arrays. At depth $d$, a sub-matrix is detected using the predicate ${\bf NOTPARTITIONED}$, and the sub-matrices are multiplied. In our analysis, we only consider nested matrices of depth 2. We refer to the first level block partitions as the node-level partitions, and the second-level partitions as the worker-level partitions.

\subsection{Elementwise Operations}
For unary operations (e.g. $-\x$), we assume $\x \in \mathbb{R}^n$ and $\x$ is partitioned into $p$ blocks. The lower bound for this operation is $\gamma p$. LSHS on Dask and the Dask scheduler will incur $0$ communication overhead. LSHS on Ray and the Ray scheduler will incur a communication overhead of approximately $R(n)$.

For binary element-wise operations, we assume $\x, \y \in \mathbb{R}^n$ and both are partitioned into $p$ blocks.
The lower bound for this operation is $\gamma p$.
LSHS on Dask achieves $0$ communication for element-wise operations: The LSHS data placement procedure ensures blocks in $\x$ and $\y$ are stored on the same workers, and the cost of executing on some other worker is strictly greater than executing on the worker on which the two input blocks already exist: The memory load is the same on whatever worker the operation is executed, but the network load is greater on workers other than the worker on which the input blocks already reside.
Similarly, LSHS on Ray ensures blocks are placed on the same nodes, but can guarantee at most $R(n)$ time due to constant overhead associated with writing function outputs to its object store.


\subsection{Reduction Operations}
Without loss of generality, we provide the communication time for $\numssum(\X)$.
Assume $\X \in \R^{n, d}$ is tall-skinny, $n >> d$, so that $\X$ is block-partitioned along its first axis into $p$ partitions of blocks with dimensions $(n/p) \times d$. For the addition operation, the $\numssum$ operation sums all blocks in $\X$ and outputs a block of $(n/p) \times d$.

Due to the same argument presented for element-wise operations, LSHS first performs $bop$ on operands which already reside on the same nodes.
Recall that $r=p/k$. The lower bound for this operation is $\gamma (p-1) + log_2(r)R(n) + log_2(k)C(n)$.

For Dask, LSHS incurs a cost of $log_2(r)D(n)$ for local reductions, plus $log_2(k)C(n)$ for the remaining $k$ blocks.
Likewise, LSHS on Ray incurs a cost of $log_2(r)R(n) + log_2(k)C(n)$.

\subsection{Block-wise Inner Product}
Assume $\X, \Y \in \R^{n, d}$ is tall-skinny, $n >> d$, so that $\X, \Y$ are block-partitioned along their first axes into $p$ partitions of blocks with dimensions $(n/p) \times d$. Under these conditions, we define the block-wise inner product as $X^\top Y$. This is the most expensive operation required to compute the Hessian matrix for generalized linear models optimized using Newton's method.

This operation will execute matrix multiplication between $X_i$ and $Y_i$, where $Z_i$ denotes the $i$th block of array $Z$. Let the output of the previous procedure be denoted by the block-partitioned array $W$. The final step of this operation is $reduce(add, W)$. Thus, the analysis provided for element-wise and reduce operations also apply to this operation.

The lower bound for this operation is $\gamma (p + p - 1) + log_2(k)C(n) + (1 + log_2(r))R(n)$. For LSHS on Dask, we have $log_2(k)C(n) + log_2(r)D(n)$, and for Ray we have $log_2(k)C(n) + (1 + log_2(r))R(n)$.

Empirically, we observe that LSHS on Ray is slightly faster than LSHS on Dask for this operation. This suggests that $(1 + log_2(r))R(n) < log_2(r)D(n)$ and $R(n) < log_2(r)(D(n) - R(n))$, which is reasonable given our assumption that $R(n) < D(n)$. As $R(n)$ goes to $0$, this inequality goes to $0 < log_2(r)D(n)$, suggesting that Dask's performance is explained by worker-to-worker communication within a single node.

\subsection{Block-wise Outer Product}
Assume $\X, \Y \in \R^{n, d}$ is tall-skinny, $n >> d$, so that $\X, \Y$ are block-partitioned along their first axes into $\sqrt{p}$ partitions of blocks with dimensions $(n/\sqrt{p}) \times d$. The block-wise outer product is defined as $\X \Y^\top$. The output $\Z$ will be a $\sqrt{p} \times \sqrt{p}$ grid of blocks. Every node must transmit $2 (\sqrt{k}-1) r C(n)$ node-level blocks to every row and column in its grid (minus itself), and every off-diagonal node must receive $2 r C(n)$ node-level blocks, resulting in a communication lower bound of $\gamma p + 2 (\sqrt{k}-1) r C(n)$, which is also the communication time attained by LSHS.

For LSHS on Dask, blocks placed within the diagonal of our $k \times k$ logical grid of nodes will not incur any inter-node communication overhead. This constitutes $k$ of the $k^2$ logical nodes. Within this diagonal grid of nodes, blocks placed within the diagonal of the $r \times r$ grid of logical workers within each node can be used to compute the output $\Z_{i,i} = \X_{i} \Y_{i}^\top$ without worker-to-worker communication. This constitutes $r$ such workers per node along the diagonal of our logical grid of nodes. We therefore incur a communication time of $k (r^2 - r)D(n)$ for blocks placed on the diagonal of the logical grid of nodes. Computing output blocks on the rest of the nodes requires an inter-node communication time of $(k^2 - k)C(n)$.


\subsection{Matrix Multiplication}
Let $\Z = \X \Y$ for $\X, \Y \in \mathbb{R}^{n \times n}$. Both are partitioned into $\sqrt{p} \times \sqrt{p}$ grids. $\Z$ will have the same dimension and partitioning as $\X$ and $\Y$.
We have $O(\sqrt{p}^3=p^{3/2})$ block operations: For each of the $\sqrt{p}^2=p$ output blocks, we have $\sqrt{p}$ matrix multiplies, and $\sqrt{p}-1$ additions.
The same arguments used to derive the communication time for block-wise inner and outer products can be applied to matrix multiplication.

Our implementation of matmul is a special case of our recursive implementation of tensordot. We can therefore view the entire computation as a $\sqrt{k} \times \sqrt{k}$ grid of node-level blocks, each of which are comprised of $\sqrt{r} \times \sqrt{r}$ worker-level blocks. At the node-level, we have $k^{3/2}$ matrix multiplies, $\sqrt{k}$ of which require $0$ inter-node communication.
The remaining $k^{3/2} - \sqrt{k}$ node-level blocks are parallelized over $k$ nodes, which requires at least $(k^{3/2} - \sqrt{k})/k r = (k-1)/\sqrt{k} r$ of the worker-level blocks to be transmitted between nodes, yielding a lower bound of $(k-1)/\sqrt{k} r C(n) < \sqrt{k} r C(n)$.

On each node, we have approximately $r \sqrt{r}$ addition operations. Over $r$ local workers, we achieve approximately $\log(\sqrt{r}) R(n)$ communication time.
At the node-level, we now need to add $k \sqrt{k}$ node-level blocks of size $rn$. With $k$ nodes, this will require approximately $\log(\sqrt{k}) r C(n)$ communication time.

Putting this all together, the communication lower bound for matrix multiplication is approximately $\left( \frac{k-1}{\sqrt{k}} + \log(\sqrt{k}) \right) r C(n) + \log(\sqrt{r}) R(n)$. We ignore the diagonal terms for the simpler expression of $\left( \sqrt{k} + \log(\sqrt{k}) \right) r C(n) + \log(\sqrt{r}) R(n)$.


\subsubsection{SUMMA}
\begin{algorithm}[ht]
\SetAlgoLined
    $\Z_{i,j} \gets {\bf 0}$;\\
    \For{$h \gets 0$ \KwTo $\sqrt{p}$}{
        Broadcast $\X_{i, h}$ to $\sqrt{p}$ workers in worker grid row $i$;\\
        Broadcast $\Y_{h, j}$ to $\sqrt{p}$ workers in worker grid column $j$;\\
        $\Z_{i,j} \gets \Z_{i,j} + \X_{i,h} \Y_{h,j}$;
    }
\caption{SUMMA}
\label{algo:summa}
\end{algorithm}

The blocked Scalable Universal Matrix Multiplication Algorithm (SUMMA) \cite{summa} is characterized by Algorithm \ref{algo:summa}. The blocks are partitioned over workers so that worker with grid coordinates $i,j$ stores blocks $\X_{i,j}$, $\Y_{i,j}$, and the output block $\Z_{i, j}$.

A tree-based broadcast has an inter-node communication cost of $(\log{\sqrt{p}}) C(n)$. The algorithm performs 2 such broadcasts per iteration over $\sqrt{p}$, yielding an inter-node communication complexity of $2 \sqrt{p} \log(\sqrt{p}) C(n)$. 

SUMMA requires $2 \sqrt{kr} \log(\sqrt{kr}) C(n) = 2\sqrt{k}(\log(\sqrt{k}) C(n) + 2\sqrt{r}\log(\sqrt{r}) C(n).$
We can see that SUMMA has a communication time of $2\sqrt{k}(\log(\sqrt{k}) C(n)$ for inter-node communication. The inter-node communication component of our lower bound is $r \left( \sqrt{k} + \log(\sqrt{k}) \right) C(n)$. We immediately see that the lower bound depends primarily on $r$. In practice, $r$ is set to the number of physical cores per node, which is relatively small ($32$ in our benchmarks). We see that, asymptotically, $2\sqrt{k}(\log(\sqrt{k})$ grows faster than $\left( \sqrt{k} + \log(\sqrt{k}) \right)$, suggesting asymptotically faster communication time in $k$. While we are unable to show that \project\ attains this lower bound, we believe it helps explain the competitive performance we achieve, especially for $k=16$ (Figure \ref{fig:matrix-multiply}). On the other hand, systems such as SLATE have no $\gamma$ term. For larger partitions, we expect $\gamma$ to be a bottleneck in systems such as \project\ and Dask Arrays.

\end{document}